\documentclass[svgnames,envcountsame,envcountsect,envcountreset]{llncs}
\usepackage{preamble} 
\usepackage[inline]{enumitem}
\usepackage{mathdots}
\usepackage{mathtools}
\usepackage{comment}
\usepackage{subcaption}
\usepackage{hyperref}
\usepackage{ifthen}
\usepackage{environ}

\newboolean{fest}
\newboolean{full}
\ifthenelse
  {\isundefined{\extendedversion}}
  {\setboolean{fest}{true}\setboolean{full}{false}}
  {\setboolean{fest}{false}\setboolean{full}{true}}

\renewenvironment{proof}[1][of \lastresult]{%
  \begin{trivlist}\item
  \emph{Proof%
  \ifthenelse{\equal{#1}{}}{}{ #1}.}
}{
  \end{trivlist}
}

\NewEnviron{fullorname}[1][\lastlabel]{%
  \iffest
     \global\expandafter\let\csname #1-lastresult\endcsname\lastresult
     \global\expandafter\let\csname #1-body\endcsname\BODY
  \else
    \BODY
  \fi
}
\newcommand{\displayhere}[1]{
  \def\lastresult{\csname#1-lastresult\endcsname}
  \csname#1-body\endcsname
}

\newcommand{\setof}[1]{\{{#1}\}}
\newcommand{\spanof}[2]{\lmangle{#1},{#2}\mrangle}

\newcommand{\spcomp}{\fatsemi}

\newcommand{\anglemidskip}{%
\mathchoice{\mskip-3.8mu}%
           {\mskip-3.8mu}%
           {\mskip-4.1mu}%
           {\mskip-4.6mu}%
}

\newcommand{\lmangle}{{\langle\anglemidskip|}}
\newcommand{\mrangle}{{|\anglemidskip\rangle}}

\newcommand{\commutes}{\circlearrowleft}

\newcommand{\st}{.\;}
\newcommand{\id}{\mathit{id}}
\newcommand{\depth}{\mathit{dp}}
\newcommand{\fv}{\mathit{fv}}

\newcommand{\sat}{\models}
\newcommand{\nsat}{\not\models}
\newcommand{\entails}{\models}

\newcommand{\ccdots}{{\cdotp\mkern-2mu\cdotp\mkern-2mu\cdotp\mkern1mu}}
\renewcommand{\to}{\rightarrow}
\newcommand{\of}{\colon}

\renewcommand{\cat}[1]{\ensuremath{\mathbf{#1}}}

\newcommand{\Rshift}{\mathbin{\nearrow_{\!\!\!\mathsf r}}}
\newcommand{\Rbshift}{\mathbin{\swarrow_{\!\!\!\mathsf r}}}
\newcommand{\Ishift}{\mathbin{\nearrow_{\!\!\mathsf i}}}
\newcommand{\Ibshift}{\mathbin{\swarrow_{\!\!\mathsf i}}}

\newcommand{\Lab}{\mathsf{L}}
\newcommand{\Var}{\mathsf{V}}
\newcommand{\FOL}{\mathsf{FOL}}

\newcommand{\ABx}{\mathsf{AB}}
\newcommand{\ACx}{\mathsf{AC}}
\newcommand{\SBx}{\mathsf{SB}}
\newcommand{\SCx}{\mathsf{SC}}

\newcommand{\AB}[1]{\ABx\if\relax\detokenize{#1}\relax\else(#1)\fi}
\newcommand{\AC}[1]{\ACx\if\relax\detokenize{#1}\relax\else(#1)\fi}

\newcommand{\SB}[1]{\SBx\if\relax\detokenize{#1}\relax\else(#1)\fi}
\newcommand{\SC}[1]{\SCx\if\relax\detokenize{#1}\relax\else(#1)\fi}

\newcommand{\bC}{\mathbf{C}}

\newcommand{\cA}{\mathcal{A}}
\newcommand{\ccA}{\cA^\circ}
\newcommand{\cB}{\mathcal{B}}
\newcommand{\cC}{\mathcal{C}}

\newcommand{\cF}{\mathcal{F}}
\newcommand{\cM}{\mathcal{S}}
\newcommand{\cN}{\mathcal{N}}
\newcommand{\cI}{\mathcal{I}}

\newcommand{\cP}{\mathcal{P}}
\renewcommand{\cR}{\mathcal{R}}
\newcommand{\cS}{\mathcal{S}}

\newcommand{\cU}{\mathcal{U}}
\newcommand{\cV}{\mathcal{V}}

\newcommand{\dF}[1]{\cF^\circ_{#1}}
\newcommand{\rF}[2]{\cF^{#2}_{#1}}
\newcommand{\dB}[1]{\cB^\circ_{#1}}
\newcommand{\rB}[2]{\cB^{#2}_{#1}}
\newcommand{\ccP}{\cP^\circ}

\newcommand{\forw}{{\mathsf{f}}}
\newcommand{\back}{{\mathsf{b}}}
\newcommand{\ashift}{{\mathsf{ash}}}
\newcommand{\sshift}{{\mathsf{ssh}}}
\newcommand{\op}[1]{{#1}^{\mathrm{op}}}

\newcommand{\ABC}{\mathbf{AC}}
\newcommand{\SBC}{\mathbf{SC}}

\newcommand{\Graph}{\cat{Graph}}

\newcommand{\Cat}{\mathbf{Cat}}
\newcommand{\SpanC}{\cat{Span}(\cat{C})}

\newcommand{\False}{\textbf{false}}
\newcommand{\True}{\textbf{true}}

\def\clap#1{\hbox to 0pt{\hss#1\hss}}

\newcommand{\gl}[1]{\mathsf{#1}}

\newcommand{\la}{\gl{a}}
\newcommand{\lb}{\gl{b}}
\newcommand{\lc}{\gl{c}}

\newcommand{\mapping}[1]{%
  \scalebox{.7}{$\begin{array}{@{}r@{\mapsto}l@{}}
  #1
  \end{array}$}
}

\newcommand{\savelabel}[1]{%
  \label{#1}%
  \gdef\lastlabel{#1}%
}
\newcommand{\dprf}{def}
\newcommand{\dname}[1]{\dprf:#1}
\newcommand{\dlabel}[1]{\savelabel{\dname{#1}}}
\newcommand{\dref}[1]{\ref{\dname{#1}}}
\newcommand{\dcite}[1]{Def.~\dref{#1}}

\newcommand{\fprf}{fig}
\newcommand{\fname}[1]{\fprf:#1}
\newcommand{\flabel}[1]{%
  \savelabel{\fname{#1}}%
}
\newcommand{\fref}[1]{\ref{\fname{#1}}}
\newcommand{\fcite}[1]{Fig.~\fref{#1}}

\newcommand{\pprf}{prop}
\newcommand{\pname}[1]{\pprf:#1}
\newcommand{\plabel}[1]{%
  \gdef\lastresult{\pcite{#1}}%
  \savelabel{\pname{#1}}%
}
\newcommand{\pref}[1]{\ref{\pname{#1}}}
\newcommand{\pcite}[1]{Prop.~\pref{#1}}

\newcommand{\thprf}{th}
\newcommand{\thname}[1]{\thprf:#1}
\newcommand{\thlabel}[1]{%
  \gdef\lastresult{\thcite{#1}}%
  \savelabel{\thname{#1}}%
}
\newcommand{\thref}[1]{\ref{\thname{#1}}}
\newcommand{\thcite}[1]{Theorem~\thref{#1}}

\newcommand{\lprf}{lem}
\newcommand{\lname}[1]{\lprf:#1}
\newcommand{\llabel}[1]{%
  \gdef\lastresult{\lcite{#1}}%
  \savelabel{\lname{#1}}
}
\newcommand{\lref}[1]{\ref{\lname{#1}}}
\newcommand{\lcite}[1]{Lemma~\lref{#1}}

\newcommand{\sprf}{sec}
\newcommand{\sname}[1]{\sprf:#1}
\newcommand{\slabel}[1]{\savelabel{\sname{#1}}}
\newcommand{\sref}[1]{\ref{\sname{#1}}}
\newcommand{\scite}[1]{Section~\sref{#1}}

\newcommand{\aprf}{app}
\newcommand{\aname}[1]{\aprf:#1}
\newcommand{\alabel}[1]{\savelabel{\aname{#1}}}
\newcommand{\aref}[1]{\ref{\aname{#1}}}
\newcommand{\acite}[1]{App.~\aref{#1}}

\newcommand{\exprf}{ex}
\newcommand{\exname}[1]{\exprf:#1}
\newcommand{\exlabel}[1]{\savelabel{\exname{#1}}}
\newcommand{\exref}[1]{\ref{\exname{#1}}}
\newcommand{\excite}[1]{Ex.~\exref{#1}}

\newtheorem{assumption}[definition]{Assumption}
\newcommand{\assprf}{ass}
\newcommand{\assname}[1]{\assprf:#1}
\newcommand{\asslabel}[1]{\savelabel{\assname{#1}}}
\newcommand{\assref}[1]{\ref{\assname{#1}}}
\newcommand{\asscite}[1]{Assumption~\assref{#1}}

\newcommand{\eqprf}{eq}
\newcommand{\eqname}[1]{\eqprf:#1}
\newcommand{\eqlabel}[1]{\savelabel{\eqname{#1}}}
\renewcommand{\eqref}[1]{(\ref{\eqname{#1}})}
\newcommand{\eqcite}[1]{Eq.~\eqref{#1}}

\usetikzlibrary{shapes.geometric}
\usetikzlibrary{positioning}
\usetikzlibrary{arrows.meta}
\usetikzlibrary{calc}

\tikzset{>=latex}
\tikzset{triangle/.style=
  {isosceles triangle,
   draw,
   shape border rotate=90,
   isosceles triangle stretches=true,
   anchor=top corner,
   minimum width=15mm,
   minimum height=15mm}}
\tikzset{graph/.style={
   inner sep=1pt,
   line width=3pt,
   draw=white,
   double=black}}
\tikzset{morphism/.style={
   color=brown!90!black,
   draw=brown!90!black,
}}
\tikzset{medge/.style={
   line width=.8pt,
   double distance=0.4pt,
   double=brown!90!black,
   draw=white,
   >/.tip={Latex[color=brown!90!black,scale=.5]}
}}
\tikzset{over/.style={
   line width=.8pt,
   double distance=0.4pt,
   double=black,
   draw=white,
   arrows={|-|},
   >/.tip={Latex[color=black,scale=.5]}
}}
\tikzset{allcolor/.style= {
   draw={#1},
   text={#1},
   >/.tip={latex[color={#1}]}
}}

\tikzset{|/.tip={Bar[black,width=.4pt,line width=2pt]}}
\tikzset{cross/.style={
   preaction={draw=white,line width=2pt}
}}

\newcommand{\tri}[6][black]{
  \path [fill=#1!10,draw=#1,opacity=.5]
        (#2)
		coordinate (#2-top)
		-- +($(#4,-#3)$)
		coordinate (#2-right)
		-- +($(#4,-#3)-(#5,0)$)
		coordinate (#2-left)
		-- (#2);
  \path (#2)
        edge[draw=none]
		node[near end,text=#1] (#2-label) {#6}
		+($(#4,-#3)-(#5/2,0)$);
}

\newcommand{\darker}[1]{#1!50!black}

\newcommand{\mygraph}[1]{%
  \renewcommand{\gl}[1]{\ensuremath{\scriptstyle{\mathsf{##1}}}}%
  \tikzset{n/.style={inner sep=1pt}}%
  \tikzset{e/.style={inner sep=1pt,pos=.4}}%
  \begin{tikzpicture}[on grid,scale=.5,inner sep=2pt,baseline=(1.south)]
  \begin{scope}[node distance=8mm]
  #1
  \end{scope}
  \end{tikzpicture}%
}
\newcommand{\inline}[1]{%
  \setlength{\fboxsep}{0pt}%
  \fbox{#1}%
}
\newcommand{\myinlinegraph}[1]{%
  \inline{\mygraph{#1}}%
}

\newcommand{\onenode}[1]{\mygraph{
\node[n] (1) {$#1$};
}}

\newcommand{\twonode}[2]{\mygraph{
\node[n] (1) {$#1$};
\node[n] (2) [right=.5 of 1] {$#2$};
}}

\newcommand{\oneloop}[2]{\mygraph{
\node[n] (1) {$#1$};
\path[e] (1) edge [->,loop right] node[right=.05] {\gl{#2}} ();
}}

\newcommand{\oneloopleft}[2]{\mygraph{
\node[n] (1) {$#1$};
\path[e] (1) edge [->,loop left] node[left=.05] {\gl{#2}} ();
}}

\newcommand{\twoloop}[4]{\mygraph{
\node[n] (1) {$#1$};
\node[n] (2) [right=of 1] {$#3$};
\path[e]
  (1) edge [->,bend left] node[near start,above] {\gl{#2}} (2)
  (2) edge [->,bend left] node[near start,below] {\gl{#4}} (1);
}}

\newcommand{\oneedge}[3]{\mygraph{
\node[n] (1) {$#1$};
\node[n] (2) [right=of 1] {$#3$};
\path[e]
  (1) edge[->] node[above] {\gl{#2}} (2);
}}

\newcommand{\oneedgeloop}[4]{\mygraph{
\node[n] (1) {$#1$};
\node[n] (2) [right=of 1] {$#3$};
\path[e]
  (1) edge[->] node[above] {\gl{#2}} (2)
  (2) edge [->,loop right] node[right=.05] {\gl{#4}} ();
}}

\newcommand{\looponeedge}[4]{\mygraph{
\node[n] (1) {$#1$};
\node[n] (2) [right=of 1] {$#4$};
\path[e]
  (1) edge[->] node[above] {\gl{#3}} (2)
  (1) edge [->,loop left] node[left=.05] {\gl{#2}} ();
}}

\newcommand{\twoedge}[5]{\mygraph{
\node[n] (1) {$#1$};
\node[n] (2) [right=of 1] {$#3$};
\node[n] (3) [right=of 2] {$#5$};
\path[e]
  (1) edge[->] node[above] {\gl{#2}} (2)
  (2) edge[->] node[above] {\gl{#4}} (3);
}}

\newcommand{\onetwoedge}[7]{\mygraph{
\node[n] (1) {$#1$};
\node[n] (2) [right=of 1] {$#3$};
\node[n] (3) [above right=.3 and 1 of 2] {$#5$};
\node[n] (4) [below right=.3 and 1 of 2] {$#7$};
\path[e]
  (1) edge[->] node[above] {\gl{#2}} (2)
  (2) edge[->] node[above] {\gl{#4}} (3)
  (2) edge[->] node[below] {\gl{#6}} (4);
}}

\newcommand{\spangraph}[5]{\mygraph{
\node[n] (1) {$#1$};
\node[n] (2) [right=of 1] {$#3$};
\node[n] (3) [right=of 2] {$#5$};
\path[e]
  (1) edge[<-] node[pos=.6,above] {\gl{#2}} (2)
  (2) edge[->] node[above] {\gl{#4}} (3);
}}

\newcommand{\cospangraph}[5]{\mygraph{
\node[n] (1) {$#1$};
\node[n] (2) [right=of 1] {$#3$};
\node[n] (3) [right=of 2] {$#5$};
\path[e]
  (1) edge[->] node[above] {\gl{#2}} (2)
  (2) edge[<-] node[pos=.6,above] {\gl{#4}} (3);
}}

\begin{document} 
\title{On Categories of Nested Conditions}

\author{Arend Rensink$^1$ \and Andrea Corradini$^2$}

\institute{
University of Twente, Netherlands,
\email{arend.rensink@utwente.nl} \and
University of Pisa, Italy,
\email{andrea.corradini@unipi.it}}
\maketitle

\begin{abstract}
Nested conditions are used, among other things, as a graphical way to express first order formulas ruling the applicability of a graph transformation rule to a given match. In this paper, we propose (for the first time) a notion of structural morphism among nested conditions, consistent with the entailment of the corresponding formulas. This reveals a structural weakness of the existing definition of nested conditions, which we overcome by proposing a new notion of \emph{span-based nested conditions}, embedding the original ones. We also introduce morphisms for the latter, showing that those form a richer structure by organising the various models in a number of categories suitably related by functors.
\end{abstract}

\section{Introduction}

Representing formulas of First-Order Logic (FOL) by graphs or more general graphical structures was explored in various areas of Theoretical Computer Science and Logics along the decades. A canonical example is represented by edge-labelled graphs, which can be regarded as an alternative syntax for formulas of a fragment of FOL including just basic (binary) predicates, equality, conjunction and existential quantification: we call this briefly the $\exists$-fragment.

Let $\Lab = \{\gl{a}, \gl{b}, \gl{c}, \ldots\}$ be a set of binary relation symbols, which we shall use also as edge labels. As an example, let $A$ be the graph $\inline{\twoedge{x}{a}{y}{b}{z}}$. We consider it as a sound representation of the formula  $\phi_A = \exists x,y,z\st \gl{a}(x,y) \land \gl{b}(y,x)$, in the following sense: a graph $G$ satisfies $\phi_A$ (or is a \emph{model} of $\phi_A$) if and only if there is a graph morphism $h$ from $A$ to $G$. 
Now consider $B = \inline{\oneedge{x}{a}{y}}$, representing formula $\phi_{B} = \exists x,y\st \gl{a}(x,y)$. Graph $B$ has an obvious inclusion morphism into $A$, viz.\ $i\of B \to A$. Therefore every morphism $h\of A \to G$ gives rise to a composed morphism $i;h\of B\to G$,\footnote{Along the paper, we denote by $f;g\of A \to C$ the composition of arrows $f\of A \to B$ and $g \of B \to C$, using diagrammatic order.} implying that every graph that satisfies $\phi_A$ also satisfies $\phi_{B}$, thus $\phi_A$ \emph{entails} $\phi_{B}$ (written $\phi_A \entails \phi_{B}$). It is worth noting that, in this elementary framework, graph morphisms can represent both a satisfaction relation (between formulas and models) and an entailment relation among formulas.

These concepts were exploited for example by Chandra and Merlin in \cite{DBLP:conf/stoc/ChandraM77} in the framework of relational database queries. They show there that every \emph{conjunctive query} (a formula of the $\exists$-fragment, but with relations of any arity) has a natural model, a graph, and query inclusion is equivalent to the existence of a graph homomorphism between those natural models. Therefore morphisms are not only sound, but also complete with respect to entailment, and it follows that query inclusion is decidable, even if NP-complete: an interesting logical result obtained with graph theoretical techniques.

In the realm of Graph Transformation Systems  (GTSs)~\cite{eept:fundamentals-agt,DBLP:books/sp/HeckelT20}
 the need of representing formulas by graphs arose in a natural way. Indeed,
in any approach a rule consists of at least two graphs, $L \leadsto R$, and to apply it to a graph $G$, first a morphism $m: L \to G$ has to be found. 
 By the above discussion, we can consider $L$ as a formula of the $\exists$-fragment that has to be satisfied by $G$, as an application condition of the rule. 

It soon turned out that in order to use GTSs for even simple specifications, more expressive application conditions were needed. 
In~\cite{NegativeAC} the authors introduced \emph{Negative Application Conditions (NACs)}, allowing to express (to some extent) negation and disjunction.  A NAC $\cN$ is  a finite set of morphisms from $L$, $\cN = \{n_i: L \to Q_i\}_{i\in[1,k]}$, and a morphism $m: L \to G$ \emph{satisfies} $\cN$ if for all $i\in[1,k]$ there is no morphism $m_i: Q_i \to G$ such that $n_i;m_i = m$. 
It follows that such a NAC represents a formula of the shape $\exists \bar{x}\st \phi_L \wedge \neg (\exists \bar{y}_1\st \phi_{Q_1} \vee \ldots \vee \exists \bar{y}_n\st \phi_{Q_n})$: as in \cite{Rensink-FOL}, we call this the $\exists \neg \exists$-fragment (of FOL).

Note that differently from the $\exists$-fragment, the structures representing the $\exists \neg \exists$-fragment are no longer graphs, but diagrams (``stars'') in \cat{Graph}, the category of graphs; and satisfaction does not require just the existence of a matching morphism from $L$, but also the non-existence of certain other morphisms.

NACs were generalized in~\cite{Rensink-FOL,Habel-FOL} to \emph{Nested (Application) Conditions}, where the structure of a condition is a finite tree of arbitrary depth rooted at $L$, and satisfaction is defined like for NACs, but iterating further at each level of the tree. Interestingly, Nested Conditions were proved to have the same expressive power of full FOL.

Since application conditions denote formulas, they are the objects of an obvious category (actually, a preorder) where arrows represent entailment. But differently from the case of the $\exists$-fragment, where entailment can be ``explained" by the existence of a (graph) morphism, we are not aware of similar results for the larger $\exists \neg\exists$-fragment or for FOL. More explicitly, despite the fact that NACs first and Nested Conditions next were defined as suitable diagrams in a category of graphs (or of similar structures), we are not aware of any definition of \emph{structural morphisms} among such application conditions, providing evidence for (some cases of) entailment like simple graph morphisms do for the $\exists$-fragment. In this paper we address exactly this issue, one main challenge being to ensure that a morphism exists between two conditions only if they are related by entailment.

We start by recalling in \scite{ab-conditions} the main definitions related to Nested Conditions from \cite{Rensink-FOL}. We call such conditions \emph{arrow-based}.
In \scite{ab-morphisms} we present the original definition of morphism among such conditions, which come in two variants giving rise to two categories of conditions. Soundness of morphisms w.r.t.~entailment is ensured by the existence of (contravariant) functors to the category of conditions with the entailment preorder.

Even if completeness of such morphisms in the sense of \cite{DBLP:conf/stoc/ChandraM77} cannot hold due to undecidability of entailment in FOL, the lack of morphisms between entailment-related conditions even in very simple cases leads us to introduce, in \scite{sb-conditions}, a variant of conditions called \emph{span-based}, for which we define in turn (in \scite{sb-morphisms}) a few notions of morphisms, and show (in \scite{categories}) that they give rise to categories into which those of arrow-based conditions embed.
In \scite{conclusion} we summarize the contributions of the paper, discuss some related work and hint at future developments.
\iffest
Full proofs of most of the statements and a few auxiliary results are collected in the appendix.

\fi
 
\section{Arrow-based conditions}
\slabel{ab-conditions}

In this section we recall the standard notion of nested condition from~\cite{Rensink-FOL}, using notations that will make the connection with the variation proposed in this paper as straightforward as possible. Here and in the remainder of the paper, we will mostly omit the term ``nested'' and just refer to \emph{conditions}; however, to distinguish between variations upon this theme, we will refer to the standard notion of nested conditions as \emph{arrow-based}.

Along the paper examples and intuitions will be based on $\cat{Graph}$, the category of directed, edge-labelled multigraphs. However, formal definitions and results will be phrased in terms of objects and arrows of a generic category $\bC$ that we assume to be a \emph{presheaf topos}, i.e., a category of contravariant functors from a small category \cat{S} to \cat{Set}, thus \cat{C} $= [\op{\cat{S}} \to \cat{Set}]$. Several categories of graphs and hypergraphs are presheaf toposes: for example, directed unlabeled graphs are obtained with \cat{S} the free category generated by $\mygraph{
  \node (1) {$\bullet$};
  \node (2) [right=of 1] {$\bullet$};
  \path (1) edge[bend left=20,->] (2)
        (1) edge[bend right=20,->] (2);
}$. Furthermore, presheaf toposes are closed under the construction of slice and functor categories, thus they include labeled/typed (hyper)graphs (see Sec.~5 of~\cite{AzziCR19}). We will denote the collection of objects of $\cat{C}$ by $|\cat{C}|$, and for $A,B \in |\cat{C}|$ we denote by $\cat{C}(A,B)$ the (hom)set of arrows from $A$ to $B$.  

Assuming that \cat{C} is a presheaf topos ensures several properties we need in the constructions of this paper: in particular, that all limits and colimits exist (and can be computed pointwise), and also that epis are stable under pullback. Furthermore, \cat{C} is \emph{adhesive}~\cite{ls:adhesive-journal}, enjoying several properties exploited in the algebraic theory of graph rewriting, where the results of this paper have potential interesting applications.\footnote{Note that requiring \cat{C} to be just adhesive would not suffice: for example, we need arbitrary pushouts, while adhesivity only guarantees pushouts along monos.}

\medskip\noindent
Arrow-based conditions are inductively defined as follows:

\begin{definition}[arrow-based condition]\dlabel{ab-condition}
  Let $R$ be an object of $\bC$. $\AC R$ (the set of \emph{arrow-based conditions} over $R$) and $\AB R$ (the set of \emph{arrow-based branches} over $R$) are the smallest sets such that
  \begin{itemize}
  \item $c\in \AC R$ if $c=(R,p_1\ccdots p_w)$ is a pair with $p_i\in \AB R$ for all $1\leq i\leq w$, where $w \geq 0$;
  \item $p\in \AB R$ if $p=(a,c)$ where $a: R\to P$ is an arrow of $\bC$ and $c\in \AC P$.
  \end{itemize}
\end{definition}
We regularly abbreviate ``arrow-based" to ``ab". We call $R$ the \emph{root} of an ab-condition or ab-branch, and $P$ the \emph{pattern} of an ab-branch (which is simultaneously the root of its subconditon). \fcite{ab-condition} provides a visualisation of an ab-condition $c$. We use $b,c$ to range over ab-conditions and $p,q$ to range over ab-branches. We use $|c|=w$ to denote the width of an ab-condition $c$, $R^c$ to denote its root, and $p^c_i=(a^c_i,c_i)$ its $i$-th branch. Finally, we use $P^c_i$ ($=R^{c_i}$) for the pattern of branch $p^c_i$. In all these cases, we may omit the superscript $c$ if it is clear from the context.
\begin{figure}[t]
\centering
\subcaptionbox
  {Condition $c=(R,p_1\ccdots p_w)$, with $p_i=(a_i,c_i)$ for $1\leq i\leq w$
   \flabel{ab-condition}}
  [.45\textwidth]
  {\begin{tikzpicture}[on grid]
  \node (Rc) {$R^c$};
  \tri{Rc}{3.0}{1.8}{3.6}{}
  \node (P1) [below left=1.6 and .7 of Rc] {$P^c_1$};
  \node [below=of Rc] {$\cdots$};
  \node (Pn) [below right=1.6 and .7 of Rc] {$P^c_w$};

  \tri{P1}{1.2}{.6}{1.2}{$c_1$}
  \tri{Pn}{1.2}{.6}{1.2}{$c_w$}
  
  \path (P1) edge[<-] node[above left] {$a^c_1$} (Rc.south)
        (Pn) edge[<-] node[above right] {$a^c_w$} (Rc.south);
\end{tikzpicture}}
\quad
\subcaptionbox
  {$g\sat c$, with responsible branch $p_i=(a_i,c_i)$ and witness $h$ such that $g=a_i;h$
   \flabel{ab-satisfaction}}
  [.5\textwidth]
  {\begin{tikzpicture}[on grid]
  \node (R) {$R$};
  \tri{R}{2.8}{1.3}{2.6}{}
  \node (Pi) [below=1.5 of R] {$P_i$};
  \tri{Pi}{1.2}{.6}{1.2}{$c_i$}
  \node (G) [right=2 of R] {$G$};

  \path (R) edge[->] node[above] {$g$} (G)
        (R) edge[->] node[left,near end] {$a_i$} (Pi)
        (Pi) edge[->,bend right] node[pos=0.2,below right] (h) {$h$} (G)
        (h) edge[draw=none] node[sloped,allow upside down] {$\nsat$} (Pi-label);
\end{tikzpicture}}
\caption{Visualisations for arrow-based conditions}
\end{figure}
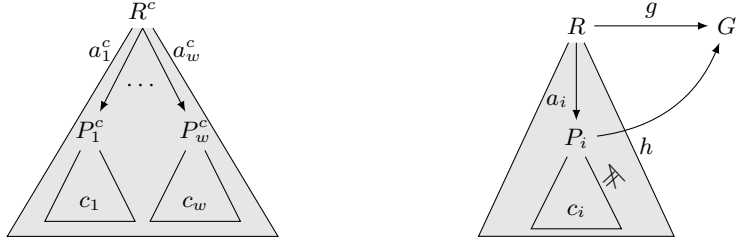

Note that, as a consequence of the inductive nature of \dcite{ab-condition}, every ab-condition has a finite \emph{depth} $\depth(c)$, defined as $0$ if $|c|=0$ and $1+\max_{1\leq i\leq |c|} \depth(c_i)$ otherwise. The depth will provide a basis for inductive proofs.

\begin{example}\exlabel{ab-conditions}
\fcite{ab-conditions} depicts three arrow-based conditions, rooted in the discrete one-node graph \inline{\onenode x}. According
to the notion of satisfaction introduced below, assuming that we already know the image of $x$ in a graph, the properties can be expressed as follows in FOL:
\begin{itemize}
\item $c_1$ is equivalent to $\lb(x,x)\wedge \nexists y\st(\la(x,y)\vee \lc(x,y))$
\item $c_2$ is equivalent to $\exists y\st \lb(x,y) \wedge \neg \la(y,y)\wedge \neg \exists z\st \lc(y,z)$ 
\item $c_3$ is equivalent to $\la(x,x)\vee (\exists y\st \lb(x,y) \wedge (\forall v,z\st \lc(y,v)\wedge \lc(y,z) \rightarrow v=z))$
\end{itemize}
Note that we have used variable names to represent nodes, to make the connection to the corresponding FOL properties more understandable. The morphisms are in all cases implied by the graph structure and variable names.
\qed
\end{example}
\begin{figure}[t]
\centering
\begin{tikzpicture}[on grid]
  \node[graph] (10) {\onenode{x}};
  \node[left=0 of 10.west,inner sep=0] {$c_1$};
  \node[graph,below=of 10] (11) {\oneloopleft{x}{b}}; 
  \node[left=0 of 11.west,inner sep=0] {$c_{11}$};
  \node[graph,below left=1 and 1.1 of 11] (111) {\looponeedge{x}{b}{a}{y}};
  \node[above right=0 of 111.north west,inner sep=1] {$c_{111}$};
  \node[graph,below right=1 and 1.1 of 11] (112) {\looponeedge{x}{b}{c}{y}};
  \node[above left=0 of 112.north east,inner sep=1] {$c_{112}$};
  
  \path (10) edge[->] (11)
        (11) edge[->] (111)
        (11) edge[->] (112);

  \node[graph] (20) [right=4.2 of 10] {\onenode{x}};
  \node[left=0 of 20.west,inner sep=0] {$c_2$};
  \node[graph,below=of 20] (21) {\oneedge{x}{b}{y}}; 
  \node[left=0 of 21.west,inner sep=0] {$c_{21}$};
  \node[graph,below left=1 and 1.1 of 21] (211) {\oneedgeloop{x}{b}{y}{a}};
  \node[above right=0 of 211.north west,inner sep=1] {$c_{211}$};
  \node[graph,below right=1 and 1.1 of 21] (212) {\twoedge{x}{b}{y}{c}{z}};
  \node[above left=0 of 212.north east,inner sep=1] {$c_{212}$};
  
  \path (20) edge[->] (21)
        (21) edge[->] (211)
        (21) edge[->] (212);
  
  \node[graph,right=4 of 20] (30) {\onenode{x}};
  \node[left=0 of 30.west,inner sep=0] {$c_3$};
  \node[graph,below left=of 30] (31) {\oneloop{x}{a}}; 
  \node[left=0 of 31.west,inner sep=0] {$c_{31}$};
  \node[graph,below right=of 30] (32) {\oneedge{x}{b}{y}}; 
  \node[right=0 of 32.east,inner sep=0] {$c_{32}$};
  \node[graph,below=1.2 of 32] (321) {\onetwoedge{x}{b}{y}{c}{v}{c}{z}}; 
  \node[above left=0 of 321.north east,inner sep=1] {$c_{321}$};
  \node[graph,below=1.2 of 321] (3211) {\twoedge{x}{b}{y}{c}{z}}; 
  \node[above left=0 of 3211.north east,inner sep=1] {$c_{3211}$};

  \path (30) edge[->] (31)
        (30) edge[->] (32)
		(32) edge[->] (321)
		(321) edge[->] node[left] {\mapping{v&z}} (3211);
\end{tikzpicture}
\caption{Examples of arrow-based conditions (see \excite{ab-conditions})}
\flabel{ab-conditions}
\end{figure}
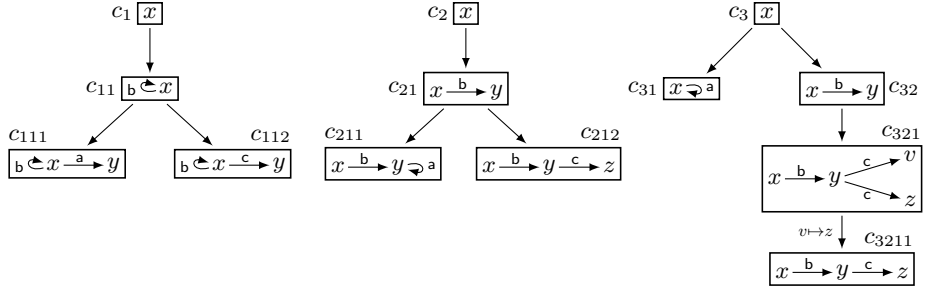

\subsection{Satisfaction}

A condition expresses a property of arrows from its root to an arbitrary object. This is operationalised through the notion of \emph{satisfaction}.

\begin{definition}[satisfaction of arrow-based conditions]\dlabel{ab-satisfaction}
  Let $c$ be an ab-condition over $R$ and $g:R\to G$ an arrow from $c$'s root to some object $G$. We say that \emph{$g$ satisfies $c$}, denoted $g\sat c$, if there is a branch $p_i=(a_i,c_i)$ of $c$ and an arrow $h:P_i\to G$ such that
  \begin{enumerate*}
  \item $g=a_i;h$ and
  \item $h\nsat c_i$.
  \end{enumerate*}
\end{definition}
If $g\sat c$, we also say that $g$ is a \emph{model} for $c$. We call $p_i$ the \emph{responsible branch} and $h$ the \emph{witness} for $g\sat c$. Pictorially, $g\sat c$ with responsible branch $p_i$ and witness $h$ can be visualised as in \fcite{ab-satisfaction}.

Based on the notion of satisfaction, for ab-conditions $b$ and $c$ (over the same root) we  define \emph{semantic entailment} $b\entails c$ and \emph{semantic equivalence} $b\equiv c$:
\begin{align*}
b \entails c & \text{ if for all arrows $g$: } g\sat b \text{ implies } g\sat c \\
b \equiv c & \text{ if for all arrows $g$: } g\sat b \text{ if and only if } g\sat c \enspace. 
\end{align*}

\begin{example}\exlabel{ab-satisfaction}
Let us consider some models for the ab-conditions in \fcite{ab-conditions}.
\begin{itemize}[topsep=0pt]
\item Let $G_1=\myinlinegraph{
\node (1) {$\bullet$};
\node (2) [right=of 1] {$\bullet$};
\path (1) edge[bend left=20,->] node[near start,above] {\lb} (2)
      (2) edge[bend left=20,->] node[near start,below] {\lb} (1)
	  (1) edge[loop left,->] node[left] {\lc} (1);
}$
and let $g$ be the morphism from the one-node discrete graph \inline{\onenode x} to the left-hand node of $G_1$. It is clear that $g\nsat c_1$ because there is no witness for $c_{11}$ ($G_1$ does not have the required $\lb$-loop). Instead, we have $g\sat c_2$: the witness $h$ (for $c_{21}$) maps $y$ to the right-hand node of $G_1$; and $h$ does not satisfy the subconditions of $c_{21}$ because it cannot be extended with either the $\la$-loop specified by $c_{211}$ or the outgoing $\lc$-edge specified by $c_{212}$. Similarly, $g\sat c_3$.

\item Let $G_2=\myinlinegraph{
\node (1) {$\bullet$};
\node (2) [right=of 1] {$\bullet$};
\node (3) [right=of 2] {$\bullet$};
\path (1) edge[bend left=20,->] node[near start,above] {\lb} (2)
      (2) edge[bend left=20,->] node[near start,below] {\lb} (1)
	  (1) edge[loop left,->] node[left] {\la} (1)
      (2) edge[->] node[above] {\lc} (3);
	  }$
and let $g,g',g''$ be the morphisms from \inline{\onenode x} to $G_2$ mapping \inline{\onenode x} to the left, mid and right node of $G_2$, respectively. None of these models satisfy $c_1$, for the same reason as above. Moreover, none satisfy $c_2$: though $g$ and $g'$ have witnesses for $c_{21}$, these are ruled out by either $c_{212}$ (in the case of $g$) or $c_{211}$ (in the case of $g'$); $g''$ not even has a witness for $c_{21}$. Instead, both $g\sat c_3$ (in fact there are two distinct witnesses, one for $c_{31}$ and one for $c_{32}$) and $g'\sat c_3$ (due to $c_{32}$); but again $g''\nsat c_3$.

\item If $G_3=\inline{\oneloopleft \bullet b}$, then the only morphism $g$ from \inline{\onenode x} to $G_3$ has $g\sat c_1$, $g\sat c_2$ and $g\sat c_3$. 
\end{itemize}
In general, it can be checked that every model of $c_1$ is also a model of $c_2$ and every model of $c_2$ is a model of $c_3$, thus $c_1\entails c_2$ and $c_2 \entails c_3$. On the other hand, $c_2\nsat c_1$ as shown by $g:\inline{\onenode x}\to G_1$, and $c_3 \nsat c_2$ as shown by $g:\inline{\onenode x}\to G_2$.\qed
\end{example}

\subsection{Connection to first-order logic}

It has been shown (e.g., \cite{Rensink-FOL,Habel-FOL}) that every first-order logic (FOL) formula can be encoded as an arrow-based nested condition over $\Graph$ and vice versa. For the sake of completeness, we summarise the connection here. We restrict to binary predicates, which we take from the set of edge labels $\Lab$. For the purpose of this discussion, w.l.o.g.\ we assume that graph nodes are variables, taken from $\Var$.

The syntax of FOL that we use is given by the grammar
\[ \phi \:::=\: \True
        \:\mid\: \False
		\:\mid\: \la(x_1,x_2)
        \:\mid\: x_1=x_2
		\:\mid\: \phi_1\wedge \phi_2
		\:\mid\: \phi_1\vee \phi_2
		\:\mid\: \neg\phi_1
		\:\mid\: \exists \bar x\st \phi_1 
		\]
where the $x_i$ are variables (from $\Var$), $\bar x\in \Var^*$ is a finite sequence of distinct variables and $\la \in \Lab$ is a binary predicate. We also use the concept of \emph{free variables} of $\phi$, denoted $\fv(\phi)$, inductively defined in the usual way.

Just as for conditions, the semantics of FOL is defined through a notion of satisfaction; however, the models are not arbitrary arrows in $\Graph$ but \emph{valuations} that map variables to the nodes of a graph. Such a valuation can be seen as an arrow from the discrete graph with node set $X\subseteq \Var$, here denoted $D_X$. Hence, the models of a formula $\phi$ are arrows $v:D_X\to G$ (for some $X\supseteq \fv(\phi)$ and some graph $G$) such that $v\sat^\FOL \phi$. For an arbitrary graph $A$ with node set $X_A$, let $v_A:D_{X_A}\to A$ map the discrete graph over $X_A$ to $A$.

Theorems 1 and~3 of \cite{Rensink-FOL} state that
\begin{enumerate*}[label=\emph{(\roman*)}]
\item for every arrow-based condition $c\in \AC R$ there exists a formula $\phi_c$ with $\fv(\phi_c)=X_R$ such that $g\sat c$ iff $v_R;g\sat^\FOL \phi_c$ for any arrow $g\of R\to G$; and

\item for every FOL-formula $\phi$, there is an arrow-based condition $c_\phi$ with $R^{c_\phi}=D_{\fv(\phi)}$ such that $v\sat^\FOL \phi$ iff $v\sat c_\phi$ for any valuation $v\of D_{\fv(\phi)}\to G$.
\end{enumerate*}
The constructions of $\phi_c$ and of $c_\phi$ are given in detail in \cite{Rensink-FOL}.

\section{Morphisms of arrow-based conditions}
\slabel{ab-morphisms}

The notion of \emph{morphism} of nested conditions has not received much attention in the literature. When considered at all, such as in \cite{bchk:conditional-reactive-systems,sksclo:coinductive-techniques-for-satisfiability}, they are essentially based on the semantics in terms of satisfaction. Indeed, entailment establishes a preorder over conditions. Let us denote the resulting category $\ABC^{\entails}$, where if $c$ and $b$ are ab-conditions, $c \leq b$ iff $c \entails b$.\footnote{Note that $c \leq b$ implies that $c$ and $b$ have the same root.} 

The question that we address in this section is to establish a meaningful \emph{structural} notion of condition morphism. That is, given the fact that an ab-condition is essentially a diagram in the category $\bC$, a structural morphism from $b$ to $c$ consists of arrows between objects of $b$ and objects of $c$ satisfying certain commutativity conditions. For morphisms to be meaningful,  they should certainly only exist where there is entailment: indeed, we require there to be a \emph{contravariant} functor to $\ABC^{\entails}$, meaning that if there is a structural morphism $b \to c$, then $c \entails b$.\footnote{This is consistent with the case of the $\exists$-fragment sketched in the Introduction.} As we will see, this requirement implies that it is not enough to consider only arrows from objects of $b$ \emph{to} objects of $c$, since in our definition of satisfaction, the direction of entailment flips at every subsequent nesting level. We also observe that the existence of a (contravariant) functor to $\ABC^{\entails}$ means that we only consider morphisms between conditions with the same root.

The underlying principle for a morphism $m$ from $c=(R,p^c_1\ccdots p^c_{|c|})$  to $b=(R,p^b_1\ccdots p^b_{|b|})$  is that it must identify, for every $b$-branch $p^b_i$, a $c$-branch $p^c_j$ that it entails. Entailment of $p^c_j$ by $p^b_i$ is captured by the existence of an arrow $v_i$ from $P^c_j$ to $P^b_i$  such that there is (recursively, but in the opposite direction with respect to $m$) a morphism between the subconditions $b_i$ and $c_j$; however, this sub-morphism can only exist ``modulo" the morphism $v_i$ between their roots. The task of the next subsection is to make the concept of ``modulo $v_i$" precise.

\subsection{Root shifting}
\slabel{shift-morphisms}

In order to relate conditions at different roots, we introduce a special kind of mapping called a \emph{root shifter}, which will later turn out to be a functor between categories of conditions over different roots. For now, we note that the purpose of root shifters is to relate the models of a condition $b\in \AC A$ to those of a condition $c\in \AC B$, given an arrow $v:A\to B$ between their roots (the existence of which implies that the model sets are typically disjoint). To be more precise, we will require that \emph{every $v$-prefixed model of $b$ induces a model for $c$}: that is, if $v;g$ is a model of $b$, then $g$ is a model of $c$.

This can be achieved in essentially two ways, both on the basis of $v$: by constructing $c$ from $b$ (which we will call \emph{forward shifting}) or by constructing $b$ from $c$ (which we will call \emph{backward shifting}).

We will first define how to shift arrows (where it is called \emph{source} shifting) and then extend this in a natural way to conditions (where it is called \emph{root} shifting).

\begin{definition}[source shifter]\dlabel{ab-source shifter}
\begin{itemize}[topsep=\smallskipamount]
\item Let us denote by $\cat{C}(A,\_)$ the set of arrows of \cat{C} with source $A$. Let $X,Y$ be objects of $\cat{C}$. A \emph{source shifter} $\cS$ from $X$ to $Y$ is a target-preserving mapping $\cS \of \cat{C}(X,\_) \to \cat{C}(Y, \_)$\footnote{That is, if $f\of X \to Z$ for some $Z$, then $\cS(f)\of Y \to Z$. More formally, $\cS$  can be defined as a family of functions $\cS = \left\{\cS_Z : \cat{C}(X, Z) \to \cat{C}(Y, Z)\right\}_{Z\in |\cat{C}|}$.} such that for all arrows $a\of X\to Z, t\of Z\to U$:
\begin{equation}\eqlabel{shifters are functors}
\cS(a;t) = \cS(a);t \enspace.
\end{equation}
\item Now let $v\of A\to B$ be an arrow. A \emph{forward source shifter $\cS$ for $v$} is a source shifter from $A$ to $B$ such that for all $a\of A\to C$, $g\of B\to G$ and $h\of C\to G$:
\begin{equation}\eqlabel{forward condition}
v;g = a;h \text{ implies } g = \cS(a);h \enspace.
\end{equation}
\item A \emph{backward source shifter $\cS$ for $v$} is a source shifter from $B$ to $A$ such that for all $a\of B\to C$, $g\of B\to G$ and $h\of C\to G$:
\begin{equation}\eqlabel{backward condition}
v;g = \cS(a);h \text{ implies } g = a;h \enspace.
\end{equation}
\end{itemize}
\end{definition}
If $\cS$ is known to be a forward shifter, we sometimes use $\cF$ to denote it; similarly, we use $\cB$ to denote backward shifters. 
Condition \eqref{shifters are functors} is actually quite strong: for $a = id_X$ it implies $\cS(id_X;t) = \cS(id_X);t$,  leading to the following observation.

%
\begin{lemma}\llabel{ab-source shifters}
Let $\cS \of \cat{C}(X,\_) \to \cat{C}(Y, \_)$  be a target-preserving mapping, and let $v\of Y\to X$ be an arrow. The following statements are equivalent:
\begin{enumerate}[topsep=\smallskipamount,label=(\alph*),ref=(\alph{enumi})]
\item\label{shifter-1} $\cS(a)=v;a$ for all $a\of X\to Z$;
\item\label{shifter-2} $\cS$ is a source shifter from $X$ to $Y$ such that $\cS(\id_X)=v$.
\end{enumerate}
\end{lemma}
\begin{fullorname}
\begin{proof}
\begin{itemize}[topsep=\smallskipamount]
\item \emph{\ref{shifter-1} implies \ref{shifter-2}.}
Assume that \ref{shifter-1} holds. To show that  $\cS$ is a source shifter, note that  \eqcite{shifters are functors} holds by $\cS(a;t)=v;a;t=\cS(a);t$. Furthermore, $\cS(\id_X)=v;\id_X=v$.

\item \emph{\ref{shifter-2} implies \ref{shifter-1}.}
Assume that \ref{shifter-2} holds. Let $a:X\to Z$; by \eqref{shifters are functors}, it follows that $\cS(a)=\cS(\id_X;a)=\cS(\id_X);a=v;a$.\qed
\end{itemize}
\end{proof}
\end{fullorname}
Knowing that source shifters from $X$ are entirely determined by $\cS(\id_X)$ gives us a basis for their construction. Before that, let us introduce some categorical terminology. An arrow $r: A \to B$ is a \emph{split epi(morphism)} if there is an arrow $m:B\to A$ such that $m;r = id_B$. In this case, arrow $m$ which satisfies the dual condition is a 
\emph{split mono(morphism)}. We also say that $r$ is a \emph{retraction} of $m$, and that $m$ is a \emph{section} of $r$.

The following proposition characterises the (forward and backward) root shifters that exist in general.
\begin{proposition}[known source shifters]\plabel{ab-source shifters}
Let $v\of A\to B$ be an arrow.
\begin{enumerate}[topsep=\smallskipamount]
\item The following three statements are equivalent.
\begin{enumerate}[label=(\alph*),ref=\theenumi(\theenumii)]
\item\label{forward-1} $\cS$ is a forward source shifter for $v$;
\item\label{forward-2} $\cS$ is a source shifter from $A$ to $B$ such that $\cS(v)=\id_B$;
\item\label{forward-3} $\cS \of \cat{C}(A,\_) \to \cat{C}(B, \_)$ is a target-preserving mapping such that $s=\cS(\id_A)$ is a section of $v$ and $\cS(a)=s;a$ for all $a\of A\to C$.
\end{enumerate}
\item If $\cS$ is a source shifter such that $\cS(\id_B)=v$, then $\cS$ is a backward source shifter for $v$ if and only if $v$ is epi.
\end{enumerate}
\end{proposition}
\begin{fullorname}
\begin{proof}
\begin{enumerate}[topsep=\smallskipamount]
\item
We show $\ref{forward-1} \Rightarrow \ref{forward-2} \Rightarrow \ref{forward-3} \Rightarrow \ref{forward-1}$.
\begin{itemize}
\item \emph{\ref{forward-1} implies \ref{forward-2}.}
Assume that \ref{forward-1} holds. By definition, $\cS$ is a source shifter from $A$ to $B$. Now consider \eqref{forward condition} with $a=v$ and $g=h=\id_B$: then $v;g=a;h$, hence $\id_B=g = \cS(a);h=\cS(a)$.

\item \emph{\ref{forward-2} implies \ref{forward-3}.}
Assume that \ref{forward-2} holds. By definition, $\cS \of \cat{C}(A,\_) \to \cat{C}(B, \_)$  is a target-preserving mapping; moreover, if $s=\cS(\id_A)$ then $s;v=\cS(\id_A);v=\cS(\id_A;v)=\cS(v)=\id_B$, proving that $s$ is a section of $v$. Moreover, $\cS(a)=\cS(\id_A;a)=\cS(\id_A);a=s;a$.

\item \emph{\ref{forward-3} implies \ref{forward-1}.}
Assume that \ref{forward-3} holds. To show \eqref{shifters are functors}, note that $\cS(a;t)=s;a;t=\cS(a);t$. To show \eqref{forward condition}, consider $a,g,h$ such that $v;g=a;h$; then $g=\id_A;g=s;v;g=s;a;h = \cS(a);h$.
\end{itemize}

\item Assume that $\cS$ is a source shifter such that $\cS(\id_B)=v$.
\begin{itemize}
\item \emph{$v$ epi implies that $\cS$ is a backward source shifter for $v$.} To show \eqref{backward condition}, let $a,g,h$ be such that $v;g=\cS(a);h=v;a;h$; then $g=a;h$ because $v$ is epi.

\item \emph{$\cS$ a backward source shifter for $v$ implies that $v$ is epi.} Let $g,h:B\to C$ be such that $v;g=v;h$, and let $a=\id_B$; then $v;g=\cS(a);h$, hence \eqref{backward condition} implies $g=a;h=\id_B;h=h$. It follows that $v$ is epi.\qed
\end{itemize}
\end{enumerate}
\end{proof}
\end{fullorname}
It follows that $v$ has precisely one forward source shifter for every section $s$ of $v$ (meaning, among other things, that $v$ must be a split epi --- hence a fortiori an epi --- to have any forward shifter at all). As for backward source shifters, if $\cS(\id_B)=v$ (which is in general the only known arrow from $A$ to $B$), then $v$ has a backward source shifter if and only if $v$ is epi. We denote these source shifters by $\cF_v^s$ and $\cB_v$, respectively. Their working is visualised in \fcite{ab-source shifters}. \fcite{ex-ab-shifters} shows some examples.

\begin{figure}[t]
\centering
\begin{tikzpicture}[on grid]
  \node (A2) {$A$};
  \node (B2) [right=2 of A2] {$B$};
  \node (B2') [below left=1 and .5 of B2] {$B$};
  \node (C2) [below right=2 and 1 of A2] {$C$};

  \path (A2) edge[->] node[left] {$a$} (C2)
        (B2') edge[->] node[right] {$\id_B$} (B2) 
        (B2') edge[->] node[right] {$\cF_v^s(a)$} (C2)
		(A2) edge[->] node [above] {$v$} (B2)
        (B2') edge[->] node[below] {$s$} (A2);
\end{tikzpicture}
\qquad
\begin{tikzpicture}[on grid]
  \node (A1) {$A$};
  \node (B1) [right=2 of A1] {$B$};
  \node (C1) [below right=2 and 1 of A1] {$C$};
  
  \path (A1) edge[->] node[left] {$\cB_v(a)$} (C1)
        (B1) edge[->] node[right] {$a$} (C1)
		(A1) edge[->>] node[above] {$v$} (B1);
\end{tikzpicture}
\caption{Forward and backward source shifting for $v:A\to B$}
\flabel{ab-source shifters}
\end{figure}
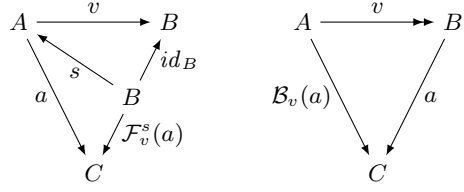

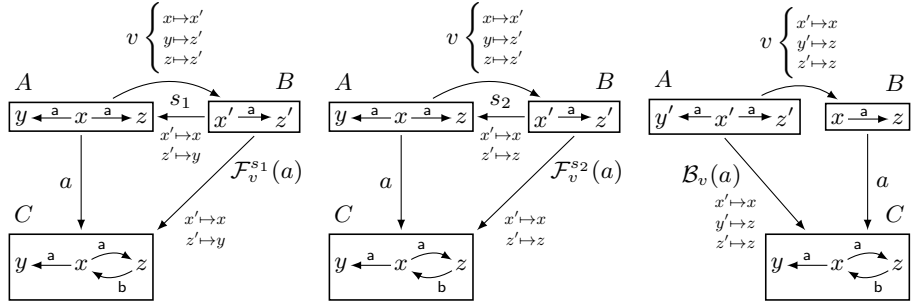
\begin{figure}
\centering
\newcommand{\Cgraph}{\mygraph{
    \node[n] (1) {$y$};
    \node[n] (2) [right=of 1] {$x$};
    \node[n] (3) [right=of 2] {$z$};
    \path (1) edge[<-] node[above] {\la} (2)
          (2) edge[->,bend left] node[near start,above] {\la} (3)
          (3) edge[->,bend left] node[near start,below] {\lb} (2);
    }
}%
\begin{tikzpicture}[on grid]
  \node[graph] (A) {\spangraph{y}{a}{x}{a}{z}};
  \node [above right=0 and 0 of A.north west] {$A$};
  \node[graph,below=2 of A] (C) {\Cgraph};
  \node [above right=0 and 0 of C.north west] {$C$};
  \node[graph] (B) [right=2.3 of A] {\oneedge{x'}{a}{z'}};
  \node [above left=0 and 0 of B.north east] {$B$};

  \path (A) edge[->] node[left] {$a$} (C);

  \path (A) edge[->,bend left] node[above]
            {$v \left\{\mapping{
			x & x' \\
			y & z' \\
			z & z'}\right.$}
	    (B);

  \path (B) edge[->] node[above] {$s_1$}
                     node[below]
            {$\mapping{
			x' & x \\
			z' & y}$}
	    (A);

  \path (B.south) edge[->] node[near start,below right,inner sep=0pt] {$\cF_v^{s_1}(a) $}  
                           node[near end,below right=-.1] 
            {$\mapping{
			x' & x \\
			z' & y}$}
		(C.north east);
\end{tikzpicture}%
\enspace
\begin{tikzpicture}[on grid]
  \node[graph] (A) {\spangraph{y}{a}{x}{a}{z}};
  \node [above right=0 and 0 of A.north west] {$A$};
  \node[graph,below=2 of A] (C) {\Cgraph};
  \node [above right=0 and 0 of C.north west] {$C$};
  \node[graph] (B) [right=2.3 of A] {\oneedge{x'}{a}{z'}};
  \node [above left=0 and 0 of B.north east] {$B$};

  \path (A) edge[->] node[left] {$a$} (C);

  \path (A) edge[->,bend left] node[above]
            {$v \left\{\mapping{
			x & x' \\
			y & z' \\
			z & z'}\right.$}
	    (B);

  \path (B) edge[->] node[above] {$s_2$} 
                     node[below]
            {$\mapping{
			x' & x \\
			z' & z}$}
	    (A);

  \path (B.south) edge[->] node[near start,below right,inner sep=0pt] {$\cF_v^{s_2}(a)$}
                           node[near end,below right=-.1] 
            {$\mapping{
			x' & x \\
			z' & z}$}
		(C.north east);
\end{tikzpicture}%
\enspace
\begin{tikzpicture}[on grid]
  \node[graph] (A) {\spangraph{y'}{a}{x'}{a}{z'}};
  \node [above right=0 and 0 of A.north west] {$A$};
  \node[graph] (B) [right=1.9 of A] {\oneedge{x}{a}{z}};
  \node [above left=0 and 0 of B.north east] {$B$};
  \node[graph,below left=2 and .4 of B] (C) {\Cgraph};
  \node [above left=0 and 0 of C.north east] {$C$};

  \path (B) edge[->] node[right] {$a$} (C.north -| B);

  \path (A) edge[->,bend left] node[above]
            {$v \left\{\mapping{
			x' & x \\
			y' & z \\
			z' & z}\right.$}
	    (B);

  \path (A.south) edge[->] node[near start,below left=-.1] {$\cB_v(a)$}
                           node[below left] 
            {$\mapping{
			x' & x \\
			y' & z \\
			z' & z}$}
		(C);
\end{tikzpicture}
\caption{Forward source shifters $\cF_v^{s_1}$ and $\cF_v^{s_2}$ and backward source shifter $\cB_v$ for an arrow $v:A\to B$. (All morphisms $a$ are the identity on nodes.)}
\flabel{ex-ab-shifters}
\end{figure}

Source shifting is extended in a straightforward manner to conditions, where it is called \emph{root shifting}. To root-shift a condition, we only have to source-shift the arrows of its branches: the subconditions remain the same. This also means that for zero-width conditions, there is essentially nothing to be done. Hence we can either extend a source shifter $\cS$ from $X$ to $Y$ to a mapping $\bar\cS\of \AC X\to \AC Y$ over conditions of arbitrary width, or use the special \emph{trivial} root shifter $\cI_{X,Y}$ for zero-width conditions, where $\bar\cS$ and $\cI_{X,Y}$ are defined as follows:
\[\begin{array}{rrl}
\bar\cS: & c & \mapsto (Y,(\cS(a^c_1),c_1)\ccdots (\cS(a^c_{|w|}),c_{|w|})) \\
\cI_{X,Y}: & (X,\epsilon) & \mapsto (Y,\epsilon) \enspace.
\end{array}\]

\begin{definition}[root shifter]\dlabel{ab-root shifter}
Let $X,Y$ be objects and $v\of A\to B$ an arrow.
\begin{itemize}[topsep=\smallskipamount]
\item A \emph{root shifter} $\cC$ from $X$ to $Y$ is a partial function from $\AC X$ to $\AC Y$ such that either $\cC=\bar\cS$ for some source shifter from $X$ to $Y$, or $\cC=\cI_{X,Y}$.

\item A \emph{forward root shifter} $\cC$ for $v$ is a root shifter from $A$ to $B$ such that either $\cC=\bar\cF$ for some forward source shifter $\cF$, or $\cC=\cI_{A,B}$.

\item A \emph{backward root shifter} $\cC$ for $v$ is a root shifter from $B$ to $A$ such that either $\cC=\bar\cB$ for some backward source shifter $\cB$, or $\cC=\cI_{B,A}$.
\end{itemize}
\end{definition}
It should be noted that \emph{all} root shifters $\bar\cS$ with $\cS$ a source shifter from $X$ to $Y$ behave like $\cI_{X,Y}$ on zero-width conditions; however, $\cI_{X,Y}$ is a root shifter for any pair of objects $(X,Y)$, whereas, as we have seen in \dcite{ab-source shifter}, forward and backward source shifters only exist for certain arrows $v\of X \to Y$. This justifies the introduction of $\cI_{X,Y}$ as a  distinguished root shifter.

\begin{fullorname}[\lname{ab-root shifter}]
The following fact, which follows from \lcite{ab-source shifters} for source shifters, will be very useful in reasoning about root shifters.
\begin{lemma}[root shifter property]\llabel{ab-root shifters}
Let $X,Y$ be objects. For any non-trivial root shifter $\cC$ from $X$ to $Y$, there is an arrow $v:Y\to X$ such that, for all $c\in \AC X$ on which $\cC$ is defined, $\cC(c)= (Y,(v;a^c_1,c_1)\ccdots (v;a^c_{|c|},c_{|c|}))$.
\end{lemma}
\begin{proof}
If $\cC$ is non-trivial, then $\cC=\bar\cS$ for some source shifter. Let $v=\cS(\id_X)$; then \lcite{ab-source shifters} implies $\cS(a)=v;a$ for any $a$ with source $A$.\qed
\end{proof}
\end{fullorname}
By ignoring from now on the notational distinction between $\cS$ and $\bar\cS$, we reuse $\cF$ and $\cB$ to range over forward and backward root shifters. The action of a root shifter is visualised in \fcite{root shifting}.
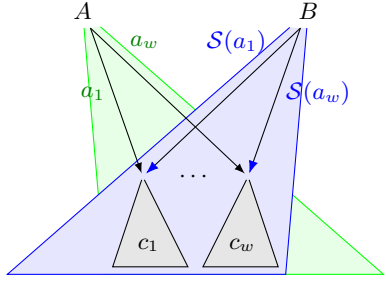
\begin{figure}[t]
\centering
\begin{tikzpicture}[on grid]
  \node (A) {$A$};
  \node (B) [right=3 of A] {$B$};
  \tri[green]{A}{3.5}{4}{3.7}{}
  \tri[blue]{B}{3.5}{-.3}{3.7}{}
  \node (P1) [below right=2.2 and .8 of A,inner sep=1] {};
  \node (Pw) [below right=2.2 and 2.2 of A,inner sep=1] {};
  \node [below right=2.2 and 1.5 of A] {$\cdots$};

  \tri{P1}{1.2}{.6}{1}{$c_1$}
  \tri{Pw}{1.2}{.4}{1}{$c_w$}
  
  \path [allcolor={\darker{green}}]
        (A.295) edge[->] node[left,pos=.45] (ab1) {$a_1$} (P1)
                edge[->] node[above right,pos=.2] (abw) {$a_w$} (Pw)
		;
  
  \path [draw=blue,text=blue,>/.tip={Latex[blue]}]
        (B.245) edge[->] node[above left,pos=.2,inner sep=1] (ac1) {$\cS(a_1)$} (P1)
                edge[->] node[right,pos=.45] (acw) {$\cS(a_w)$} (Pw)
		;
\end{tikzpicture}
\caption{Root shifting using a source shifter $\cS$: green is shifted to blue}
\flabel{root shifting}
\end{figure}
The following is the crucial preservation property of root shifters:
\begin{proposition}[root shifters preserve models]\plabel{ab-root shifters preserve}
Let $v:A\to B$.
\begin{enumerate}[topsep=\smallskipamount]
\item If $c\in \AC A$ and $\cF$ is a forward root shifter for $v$ that is defined on $c$, then $v;g\sat c$ implies $g\sat \cF(c)$.
\item If $c\in \AC B$ and $\cB$ is a backward root shifter for $v$ that is defined on $c$, then $v;g\sat \cB(c)$ implies $g\sat c$.
\end{enumerate}
\end{proposition}
\begin{fullorname}
\begin{proof}
For zero-width conditions, both clauses are vacuously true. In the remainder assume $|c|>0$ (and hence the root shifters are non-trivial). Let $v:A\to B$.
\begin{enumerate}
\item Let $b=\cF(c)$; then $a^b_i=\cF(a^c_i)$ for all $1\leq i\leq |c|$. Assume $v;g\sat c$ due to responsible branch $p^c_i$ and witness $h$; i.e., $v;g=a^c_i;h$. \eqcite{forward condition} then implies $g=\cF(a);h$, hence $p^b_i$ is a responsible branch and $h$ a witness for $g\sat b$.
\item Let $b=\cB(c)$; then $a^b_i=\cB(a^c_i)$ for all $1\leq i\leq |c|$. Assume $v;g\sat b$ due to responsible branch $p^b_i$ and witness $h$; i.e., $v;g=\cB(a^c_i);h$. \eqcite{backward condition} then implies $g=a;h$, hence $p^c_i$ is a responsible branch and $h$ a witness for $g\sat c$.
\qed
\end{enumerate}
\end{proof}
\end{fullorname}
Source and root shifters compose, in the following way.
\begin{proposition}[identities and composition of shifters]\plabel{ab-shifters compose}
\begin{enumerate}[topsep=\smallskipamount]
\item For every $X$, the identity function on $\cat{C}(X,\_)$, denoted $\cI_X$ and called the identity shifter, is a source shifter that is forward and backward for $\id_X$.
\item If $\cU$ is a source [root] shifter from $X$ to $Y$ and $\cV$ a source [root] shifter from $Y$ to $Z$, then $\cU;\cV$ (understood as partial function composition) is a source [root] shifter from $X$ to $Z$. 
\item If $\cU$ and $\cV$ are forward source [root] shifters for $u: X\to Y$ and $v: Y \to Z$, respectively, then $\cU;\cV$ is a forward source [root] shifter for $u;v$.
\item Dually, if $\cU$ and $\cV$ are backward source [root] shifters for $u: Y\to X$ and $v: Z \to Y$, respectively, then $\cU;\cV$ is a backward source [root] shifter for $v;u$.
\end{enumerate}
\end{proposition}

\begin{fullorname}
\begin{proof}
We first prove the case for source shifters.
\begin{enumerate}
\item To prove \eqcite{shifters are functors} for $\cU;\cV$ consider:
\[ (\cU;\cV)(a;t) = \cV(\cU(a;t))= \cV(\cU(a);t)=\cV(\cU(a));t=(\cU;\cV)(a);t \enspace. \]

\item In case $\cU$ and $\cV$ are forward, to prove \eqcite{forward condition} assume $u;v;g=a;h$. By applying \eqcite{forward condition} for $\cU$ and $\cV$ it follows that $v;g=\cU(a);h$ and hence $g=\cV(\cU(a));h=(\cU;\cV)(a)$.

\item In case $\cU$ and $\cV$ are backward, to prove \eqcite{backward condition} for $\cU;\cV$ assume $v;u;g=(\cU;\cV)(a);h=\cV(\cU(a));h$. By applying \eqcite{backward condition} for $\cV$ and $\cU$ it follows that $u;g=\cU(a);h$ and hence $g=a;h$.
\end{enumerate}
Now assume $\cU$ and $\cV$ are root shifters. If both are non-trivial, then so is $\cU;\cV$ and the result follows by a straightforward pointwise argument. If, on the other hand, either $\cU$ or $\cV$ is trivial, then in fact so is $\cU;\cV$, which immediately establishes the result.
\qed
\end{proof}
\end{fullorname}

\subsection{Structural morphisms}

The following defines structural morphisms, based on root shifters, over ab-conditions with the same root.

\begin{definition}[arrow-based condition morphism]\dlabel{ab-morphism}
  Let $c,b \in \AC{R}$. A \emph{forward-shift [backward-shift] condition morphism} $m: c \to b$ is a pair $(o,(v_1,m_1)\ccdots (v_{|b|},m_{|b|}))$ where
  \begin{itemize}[topsep=\smallskipamount]
  \item $o:[1,|b|]\to[1,|c|]$ is a function from $b$'s branches to $c$'s branches;
  \item for all $1\leq i\leq |b|$, $v_i:P^c_{o(i)}\to P^b_i$ is an arrow from the pattern of $p^c_{o(i)}$ to that of $p^b_i$ such that $a^c_{o(i)};v_i=a^b_i$;
  \item\emph{Forward shift}: for all $1\leq i\leq |b|$, there is a forward root shifter $\cF_i$ for $v_i$ such that $m_i:b_i \to \cF_i(c_{o(i)})$ is a forward-shift morphism.
  \item\emph{Backward shift}: for all $1\leq i\leq |b|$, there is a backward root shifter $\cB_i$ for $v_i$ such that $m_i:\cB_i(b_i) \to c_{o(i)}$ is a backward-shift morphism.
  \end{itemize}
\end{definition}
We use the same notational conventions as for conditions: $m$ has width $|m|$ and depth $\depth(m)$, and the components of a morphism $m$ are denoted $o^m$, $v^m_i$ and $m_i$ for all $1\leq i\leq |m|$. \fcite{shifted-morphism} visualises the principle of forward- and backward-shift morphisms.
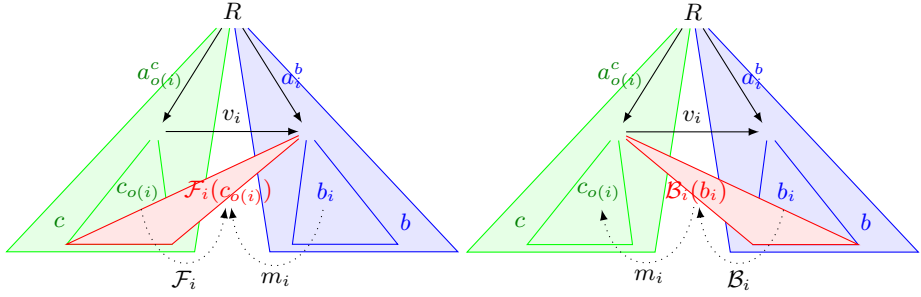
\begin{figure}[t]
\centering
\begin{tikzpicture}[on grid]
\node (R) {$R$};
\tri[green]{R}{3.2}{-.5}{2.5}{}
\node [below left=2.8 and 2.3 of R,text=\darker{green}] {$c$};
\node (Pci) [below left=1.6 and 1.0 of R] {};
\path [allcolor={\darker{green}}] (R) edge[->] node[left] {$a^c_{o(i)}$} (Pci);
\tri[green]{Pci}{1.5}{.2}{1.4}{}
\node (ci) [below left=.8 and .25 of Pci,text=\darker{green}] {$c_{o(i)}$};

\tri[blue]{R}{3.2}{3.0}{2.5}{}
\node [below right=2.8 and 2.3 of R,text=blue] {$b$};
\node (Pbi) [below right=1.6 and 1.0 of R] {};
\path [allcolor=blue] (R) edge[->] node[right] {$a^b_i$} (Pbi);
\tri[blue]{Pbi}{1.5}{1.2}{1.4}{}
\node (bi) [below right=.8 and .25 of Pbi,text=blue] {$b_i$};

\path (Pci) edge[->] node[above] {$v_i$} (Pbi);

\tri[red]{Pbi}{1.5}{-1.8}{1.4}{}
\node [left=1.3 of bi,red,inner sep=1pt] (Bi) {$\cF_i(c_{o(i)})$};

\path [->,bend left=80,below,dotted]
      (bi) edge[looseness=2.1] node {$m_i$} (Bi)
      (ci) edge[looseness=2.2,bend right=80] node {$\cF_i$} (Bi)
	  ;
\end{tikzpicture}
%
%
\begin{tikzpicture}[on grid]
\node (R) {$R$};
\tri[green]{R}{3.2}{-.5}{2.5}{}
\node [below left=2.8 and 2.3 of R,text=\darker{green}] {$c$};
\node (Pci) [below left=1.6 and 1.0 of R] {};
\path [allcolor={\darker{green}}] (R) edge[->] node[left] {$a^c_{o(i)}$} (Pci);
\tri[green]{Pci}{1.5}{.2}{1.4}{}
\node (ci) [below left=.8 and .25 of Pci,text=\darker{green}] {$c_{o(i)}$};

\tri[blue]{R}{3.2}{3.0}{2.5}{}
\node [below right=2.8 and 2.3 of R,text=blue] {$b$};
\node (Pbi) [below right=1.6 and 1.0 of R] {};
\path [allcolor=blue] (R) edge[->] node[right] {$a^b_i$} (Pbi);
\tri[blue]{Pbi}{1.5}{1.2}{1.4}{}
\node (bi) [below right=.8 and .25 of Pbi,text=blue] {$b_i$};

\path (Pci) edge[->] node[above] {$v_i$} (Pbi);

\tri[red]{Pci}{1.5}{3.2}{1.4}{}
\node [left=1.2 of bi,red,inner sep=1pt] (Bi) {$\cB_i(b_i)$};

\path [->,bend left=80,below,dotted]
      (bi) edge[looseness=2.2,bend left=80] node {$\cB_i$} (Bi)
      (Bi) edge[looseness=2.0,bend left=80] node {$m_i$} (ci)
	  ;
\end{tikzpicture}
\caption{Visualisation of forward-shift and backward-shift morphisms $m:c \to b$}
\flabel{shifted-morphism}
\end{figure}
The next result states that the existence of a morphism between two conditions $m: c \to b$ provides evidence that $b\entails c$.

\begin{proposition}[ab-condition morphisms reflect models]
  \plabel{ab-morphisms preserve models}
Let $c,b \in \AC{R}$  be arrow-based conditions. If $m:c\to b$ is a forward-shift or a backward-shift morphism, then $g\sat b$ implies $g \sat c$ for all arrows $g:R\to G$. 
\end{proposition}
\begin{fullorname}
\begin{proof}
Let $m = (o,(v_1,m_1)\ccdots (v_{|b|},m_{|b|}))$ and assume by induction hypothesis that for all $1\leq j\leq |b|$ the property holds for $m_j$. Let $p^b_i$ be the responsible branch and $h: P^b_i \to G$ the witness for $g\sat b$. 
Hence we have $(\dagger)\, g=a^b_i;h$ and $(\ddagger)\, h \nsat b_i$.  
We show that $g \sat c$, with responsible branch $p^c_{o(i)}$ and witness $h$. In fact, we immediately have  $a^c_{o(i)}; v_i ;h = a^b_i ; h \stackrel{(\dagger)}{=} g$. It remains to show that $v_i ;h \nsat c_{o(i)}$, for which we consider separately the two cases.
\begin{enumerate}
\item Let $m$ be a forward-shift morphism. Then for all $1\leq j\leq |b|$ we have that   $m_j:b_j \to \cF_j(c_{o(j)})$ is a forward-shift morphism, where $\cF_j$ is a forward root shifter for $v_j$. If (ad absurdum) $v_i ;h \sat c_{o(i)}$, by clause 1 of \pcite{ab-root shifters preserve} we would have $h \sat \cF_i(c_{o(i)})$, and by induction hypothesis on $m_i$ also $h \sat b_i$, contradicting $(\ddagger)$.

\item Let $m$ be a backward-shift morphism. Then for all $1\leq j\leq |b|$ we have that  $m_j:\cB_j(b_j) \to c_{o(j)}$ is a backward-shift morphism, where $\cB_j$ is a backward root shifter of width $|b_j|$.
Suppose (ad absurdum) that $v_i ;h \sat c_{o(i)}$. Then by the induction hypothesis on $m_i$ we also have $v_i ;h \sat \cB_i(b_i)$, and by clause 2 of \pcite{ab-root shifters preserve} we  have $h \sat b_i$, contradicting $(\ddagger)$.
\qed
\end{enumerate}
\end{proof}
\end{fullorname}

\begin{example}\exlabel{ab-morphisms}
Going back to \fcite{ab-conditions}, though (as noted before) $c_1\entails c_2$, there does not exist a morphism $m\of c_2\to c_1$. A mapping $v_1$ does exist from $c_{21}$ to $c_{11}$; however, no mapping $v_{11}$ can be found to $c_{211}$ from either $c_{111}$ or $c_{112}$. 

On the other hand, $c_2\entails c_3$ has evidence in the form of a morphism $m\of c_3\to c_2$, with the identity mapping $v_1\of c_{32} \to c_{21}$ (in combination with either the forward shifter $\cF_{v_1}^{v_1}$ or the backward shifter $\cB_{v_1}$) and a corresponding sub-morphism $m_1$ with mapping $v_{12}:c_{212}\to c_{321}$ (in combination with the trivial root shifter).
\qed
\end{example}
\pcite{ab-morphisms preserve models} can be phrased more abstractly as the existence of a contravariant functor from a category of arrow-based conditions and their (forward-shift or backward-shift) morphisms to $\ABC^{\entails}$. An important step to arrive at that result is to show that morphisms compose. This, in turn, depends on the preservation of morphisms by root shifters.

\begin{lemma}[root shifters preserve morphisms]
             \llabel{ab-root shifters preserve morphisms}
Let $c,b \in  \AC{X}$ and let $\cC$ be a root shifter from $X$ to $Y$ that is defined on $c$ and $b$. If $m:c\to b$, then also $m:\cC(c)\to\cC(b)$. 
\end{lemma}
\begin{fullorname}
\begin{proof}
If $\cC$ is trivial, the result is immediate. Otherwise, assume $m = (o,(v_1,m_1)\cdots(v_{|b|},m_{|b|}))$ and let $v$ be as in \lcite{ab-root shifters}. The only new proof obligation for $m$ to be an ab-morphism from $c$ to $b$ is that, for all $1\leq i\leq |b|$, $v;a^c_{o(i)};v_i=v;a^b_i$. This follows immediately from $a^c_{o(i)};v_i=a^b_i$.\qed
\end{proof}
\end{fullorname}
Composition of ab-condition morphisms can be defined inductively, and is independent of their forward or backward nature. We also (inductively) define arrow-based identity morphisms. Let $e,c,b$ be ab-conditions and $m:e\to c,n:c\to b$ ab-morphisms, with $|b|=w$. Then $\id_b$ and $m;n$ are defined as follows: 
\begin{eqnarray}
\id_b & =
  & (\id_{[1,w]},(\id_{P^b_1},\id_{b_1})\cdots 
                 (\id_{P^b_w},\id_{b_w}))
  \eqlabel{id-morphism} \\
m;n & =
  & (o^n;o^m,(v^m_{o^n(1)};v^n_1,n_1;m_{o^n(1)})\cdots 
              (v^m_{o^n(w)};v^n_w,n_w;m_{o^n(w)})) \enspace.
 \eqlabel{morphism composition}
\end{eqnarray}
Identities are morphisms, and morphism composition is well-defined.

\begin{lemma}[identities and composition of ab-morphisms]\llabel{ab-morphisms compose}
Let $e,c,b$ be ab-conditions and $m\of e\to c, n\of c\to b$ ab-condition morphisms.
\begin{enumerate}[topsep=\smallskipamount]
\item $\id_b$ is both a forward- and a backward-shift morphism from $b$ to $b$;
\item If $m$ and $n$ are forward-shift morphisms, then $m;n$ is a forward-shift morphism from $e$ to $b$;
\item If $m$ and $n$ are backward-shift morphisms, then $m;n$ is a backward-shift morphism from $e$ to $b$.
\end{enumerate}
\end{lemma}
\begin{fullorname}
\begin{proof}
Let $w=|b|$.
\begin{enumerate}[topsep=\smallskipamount]
\item By induction on the depth of $b$. For all $1\leq i\leq w$, $a^b_i;\id_{P^b_i}=a^b_i$. Note that $\rF{\id}{\id}$ and $\cB_\id$ (with $\id=\id_{P^b_i}$) are, respectively, forward and backward root shifters, both acting as the identity on arrows with source $P^b_i$; hence (by the induction hypothesis) $\id_{b_i}$ is both a forward-shift morphism from $b_i$ to $\rF\id\id(b_i)$ and a backward-shift morphism from  $\cB_\id(b_i)$ to $b_i$.
\end{enumerate}
The other two clauses are proved by induction on the depth of $m$ and $n$. Let $o=o^n;o^m$ and for all $1\leq i\leq w$ let $v_i=v^m_{o^n(i)};v^n_i$. Note that $o$ is a function from $[1,w]$ to $[1,|e|]$ and all $v_i$ are arrows from $P^e_{o(i)}$ to $P^b_i$ as required.
\begin{enumerate}[resume]
\item For all $1\leq i\leq |b|$ and $1\leq j\leq |c|$, let $\cF^n_i$ and $\cF^m_j$ be the foward root shifters for $v^n_i$ and $v^m_j$, respectively, guaranteed by the fact that $m$ and $n$ are forward-shift morphisms, i.e., such that $n_i:b_i \to \cF^n_i(c_{o^n(i)})$ and $m_j:c_j \to \cF^m_j(e_{o^m(j)})$ are forward-shift morphisms.

\smallskip
For all $1\leq i\leq w$, we need to show the existence of a forward root shifter $\cF_i$ for $v_i$ such that $n_i;m_{o^n(i)}$ is a morphism from $b_i$ to $\cF_i(e_{o(i)})$. Let $j=o^n(i)$ (hence $o^m(j)=o(i)$) and let $\cF_i=\cF^m_j;\cF^n_i$. By \pcite{ab-shifters compose}, $\cF_i$ is a forward root shifter for $v^m_j;v^n_i=v_i$. By definition, $m_j$ is a forward-shift morphism from $c_j$ to $\cF^m_j(e_{o^m(j)})$ and hence by \lcite{ab-root shifters preserve morphisms} (applying $\cF^n_i$) also a forward-shift morphism from $\cF^n_i(c_j)$ to $\cF^n_i(\cF^m_j(e_{o^m(j)}))$. Again by definition, $n_i$ is a forward-shift morphism from $b_i$ to $\cF^n_i(c_j)$. By the induction hypothesis, therefore, $n_i;m_j$ is a forward-shift morphism from  $b_i$ to $\cF_i(e_{o(i)})$.

\item Mutatis mutandis.
\qed
\end{enumerate}
\end{proof}
\end{fullorname}
We now state the first main result: ab-conditions endowed with either forward or backward morphisms give rise to a category with a contravariant functor to $\ABC^{\entails}$.

\begin{theorem}[categories of arrow-based conditions]\thlabel{ab-categories}
Let $\ABC^{\forw}$ be the category having ab-conditions as objects and forward-shift morphisms as arrows, and let $\ABC^{\back}$ be the category having the same objects and backward-shift morphisms as arrows. Both categories are well-defined, and there are identity-on-objects functors $\op{(\ABC^{\forw})} \to \ABC^{\entails}$ and $\op{(\ABC^{\back})} \to \ABC^{\entails}$.
\end{theorem}
\emph{Proof sketch.} Composition of morphisms and the fact that all identity morphisms have the correct nature for both categories is shown by \lcite{ab-morphisms compose}.
Moreover, identity morphisms and morphism composition satisfy the unit and associativity laws. The existence of the two identity-on-object functors follows from \pcite{ab-morphisms preserve models}.
\qed 

\medskip\noindent
Even though both forward- and backward-shift morphisms preserve models, they are incomparable: if two conditions are related by one kind of morphisms they are not necessarily related by the other kind. In other words, if we consider the existence of a morphism as providing evidence for entailment, then these two types of morphism are complementary, as the following example shows.
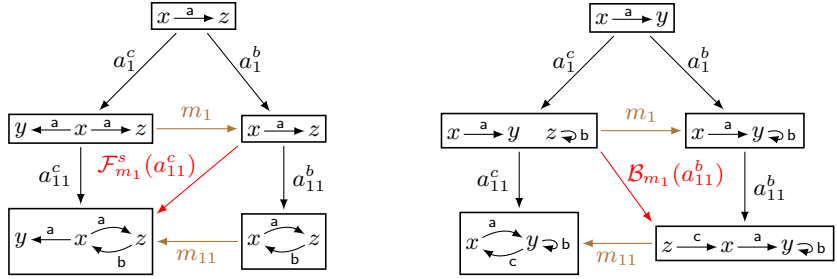
\begin{figure}[t]
\centering
\subcaptionbox
  {A forward-shift morphism $m \of c \to b$ without a backward-shift one
   \flabel{forward-no-backward}
  }
  [.45\textwidth]
  {\begin{tikzpicture}[on grid]
  \node[graph] (R1) {\oneedge x a z};
  \node[graph,below right=1.5 and 1.2 of R1] (10) {\oneedge{x}{a}{z}};
  \node[graph,below=1.5 of 10] (11) {\twoloop{x}{a}{z}{b}}; 

  \path (R1) edge[->] node[above right,inner sep=0] {$a^b_1$} (10)
        (10) edge[->] node[right] {$a^b_{11}$} (11);

  \node[graph,below left=1.5 and 1.5 of R1] (20) {\spangraph{y}{a}{x}{a}{z}};
  \node[graph,below=1.5 of 20] (21) {\mygraph{
    \node[n] (1) {$y$};
    \node[n] (2) [right=of 1] {$x$};
    \node[n] (3) [right=of 2] {$z$};
    \path (1) edge[<-] node[above] {\la} (2)
          (2) edge[->,bend left] node[near start,above] {\la} (3)
          (3) edge[->,bend left] node[near start,below] {\lb} (2);
    }};

  \path (R1) edge[->] node[above left,inner sep=0] {$a^c_1$} (20)
        (20) edge[->] node[left] {$a^c_{11}$} (21);
  
  \path[morphism]
        (20) edge[->] node[above] {$m_1$} (10)
        (11) edge[->] node[below] {$m_{11}$} (21);
	
\path[red,color=red]
  (10.south west) edge[->] node[above left,inner sep=0pt] {$\cF_{m_1}^s(a^c_{11})$} (21.20);
\end{tikzpicture}}
  \qquad
\subcaptionbox
  {A backward-shift morphism $m \of c \to b$ without a forward-shift one
   \flabel{backward-no-forward}
  }
  [.45\textwidth]
  {\begin{tikzpicture}[on grid]
  \node[graph] (R) {\oneedge x a y};
  \node[graph,below left=1.5 and 1.5 of R] (20) {\mygraph{
    \node[n] (1) {$x$};
    \node[n] (2) [right=of 1] {$y$};
    \node[n] (3) [right=.5 of 2] {$z$};
    \path (1) edge[->] node[above] {\la} (2)
          (3) edge[loop right,->] node[right] {\lb} (3);
    }};
  \node[graph,below=1.5 of 20] (21) {\mygraph{
    \node[n] (1) {$x$};
    \node[n] (2) [right=of 1] {$y$};
    \path (1) edge[->,bend left] node[near start,above] {\la} (2)
          (2) edge[->,bend left] node[near start,below] {\lc} (1)
          (2) edge[loop right,->] node[right] {\lb} (2);
    }};

  \path (R) edge[->] node[above left,inner sep=0] {$a^c_1$} (20)
        (20) edge[->] node[left] {$a^c_{11}$} (21);
  
  \node[graph,below right=1.5 and 1.5 of R] (10) {\mygraph{
    \node[n] (1) {$x$};
    \node[n] (2) [right=of 1] {$y$};
    \path (1) edge[->] node[above] {\la} (2)
          (2) edge[loop right,->] node[right] {\lb} (2);
    }};
  \node[graph,below=1.5 of 10] (11) {\mygraph{
    \node[n] (1) {$x$};
    \node[n] (2) [right=of 1] {$y$};
    \node[n] (3) [left=of 1] {$z$};
    \path (1) edge[->] node[above] {\la} (2)
          (3) edge[->] node[above] {\lc} (1)
          (2) edge[loop right,->] node[right] {\lb} (2);
    }}; 

  \path (R) edge[->] node[above right,inner sep=0] {$a^b_1$} (10)
        (10) edge[->] node[right] {$a^b_{11}$} (11);

  \path[morphism]
        (20) edge[->] node[above] {$m_1$} (10)
        (11) edge[->] node[below] {$m_{11}$} (21);
		
\path[red,color=red]
  (20.south east) edge[->] node[above right,inner sep=0pt] {$\cB_{m_1}(a^b_{11})$} (11.north west);
\end{tikzpicture}}
\caption{Forward-shift versus backward-shift morphisms}
\flabel{forward vs backward}
\end{figure}
\begin{example}[forward versus backward root shifters]\exlabel{forward vs backward}
\fcite{forward-no-backward} shows a forward-shift morphism $m \of c \to b$. 
Both conditions are evaluated in a context where we know $\la(x,z)$. In that context,  $c$ is equivalent to $\exists y\st \la(x,y) \wedge \neg \lb(z,x)$, and $b$ is equivalent to the property $\neg \lb(z,x)$. Intuitively, $b\entails c$ holds because $y$ can be instantiated with $z$. The figure shows $\cF_{m_1}^s(a^c_{11})$ (for a section $s$ of $m_1$), which acts as the identity on node and edges. Note that there is no backward-shift morphism $m'\of c \to b$ because $\cB_{m_1}(a^b_{11}) ; m_{11} = m_1; a^b_{11} ; m_{11} \not = a^c_{11}$.

\fcite{backward-no-forward} shows a backward-shift morphism $m \of c \to b$. 
Both conditions are evaluated in a context where we know $\la(x,y)$. In that context, $c$ is equivalent to $\exists z.\lb(z,z) \wedge \neg(y=z \wedge \lc(y,x))$ and $b$ is equivalent to $\lb(y,y) \wedge \nexists z\st \lc(z,x)$.
Indeed, $b\entails c$ because $z$ can be instantiated with the value of $y$. The figure shows $\cB_{m_1}(a^b_{11})$. Note that there cannot be a forward-shift morphism $m'\of c \to b$ because $m_1$ is not split epi.
\qed
\end{example}
As a final observation, we can characterise the relation between shifters and conditions as \emph{indexed categories}. For this purpose, let $\bC^\ashift$ denote the category with objects from $\bC$ (which as usual can be any presheaf topos) and arrows $\cS\of X\to Y$ whenever $\cS$ is an (arrow-based) source shifter from $Y$ to $X$. Identity and composition are defined as in \pcite{ab-shifters compose}.

Moreover, for every $\bC^\ashift$-object $X$ let $\ACx^\forw(X)=\ABC^\forw(X)$ (the full subcategory of $\ABC^\forw$ containing only conditions rooted at $X$), and for every $\bC^\ashift$-arrow $\cS\of X\to Y$ let $\ACx^\forw(\cS)$ be the functor that maps every $c\in \AC Y$ to $\bar\cS(c)\in \AC X$ and every forward-shift morphism $m\of c\to b$ (for $c,b\in \AC Y$) to $m\of \bar\cS(c)\to \bar\cS(b)$. Define $\ACx^\back$ analogously for backward-shift morphisms.

\begin{proposition}[$\bC^\ashift$-indexed categories]\plabel{ab indexed}
$\ACx^\forw$ and $\ACx^\back$ are functors from $\op{(\bC^\ashift)}$ to $\Cat$, establishing $\bC^\ashift$-indexed categories.
\end{proposition}
Note that this is only correct because (implicitly) $\bC^\ashift$ is a well-defined category, and for all $\cS\of X\to Y$, $\ACx^\forw(\cS)$ and $\ACx^\back(\cS)$ are well-defined functors from $\ABC^\forw(Y)$ to $\ABC^\forw(X)$ and from $\ABC^\back(Y)$ to $\ABC^\back(X)$, respectively.

\section{Span-based conditions}
\slabel{sb-conditions}

The peculiar definition of morphisms of ab-conditions just introduced, with arrows changing direction at each layer in order to ensure the preservation of models, clashes with a structural property of arrow-based conditions. Indeed, the pattern of each subcondition must contain a (homomorphic) copy of the parent pattern, because a branch consists of an arrow of $\cat{C}$ from the latter to the former. Thus if a morphism maps, say, the pattern $P^c$ of (a subcondition of) $c$ to a pattern $P^b$ of $b$, at the next layer (the copy of) $P^b$ must be mapped backwards to (the copy) of $P^c$. The consequence is that morphisms fail to exist even in very simple situations. An example is given by \excite{ab-morphisms}, where we have seen that $c_1\entails c_2$ (see \fcite{ab-conditions}) but there is no morphism providing evidence for this.
  
Inspired by this observation, we present the second main contribution of this paper: an original definition of nested conditions where the duplication of structure just described does not occur, because at each level we need to specify only the \emph{additional} structure, and how it is connected. This is achieved by replacing the branch arrows $a^c_i$ of a condition $c$ by \emph{spans}.

A span $s = (f: I \to A, g:I \to B)$ over \cat{C}, which we will denote $\spanof f g$, is a pair of arrows of \cat{C} having the same source. Objects $A$ and $B$ are the \emph{source} and \emph{target} of $s$, respectively, while $I$ is its \emph{interface}; we sometimes write $s:A\to B$. Spans can be composed: if the target of $\spanof u d$ is equal to the source of $\spanof v w$, then $\spanof u d \spcomp \spanof v w=\spanof{v';u}{d';w}$ where $\spanof {v'}{d'}$ is the pullback span of $(d,v)$. The result is defined, because \cat{C} has all pullbacks, but non-deterministic in general because pullbacks are defined up to iso. The standard solution would be to consider spans up to isomorphism of their interface (yielding category {\SpanC} having the same objects of \cat{C} and equivalence classes of spans as arrows), but to avoid further technicalities we will gloss over this throughout the paper.

\medskip\noindent
Span-based conditions are inductively defined as follows:

\begin{definition}[span-based condition]\dlabel{sb-condition}
  For any object $R$ of $\bC$, $\SC R$ (the set of \emph{span-based conditions} over $R$) and $\SB R$ (the set of \emph{span-based branches} over $R$) are the smallest sets such that
  \begin{itemize}
  \item $c\in \SC R$ if $c=(R,p_1\ccdots p_w)$ is a pair with $p_i\in \SB R$ for all $1\leq i\leq w$;
  \item $p\in \SB R$ if $p=(\spanof u d,c)$ where $u: I\to R, d:I\to P$ form a span of arrows of $\bC$ and $c\in \SC P$.
  \end{itemize}
\end{definition}
In the span $\spanof u d$ of a branch $p=(\spanof u d,c)$, $u$ stands for the \emph{up-arrow} and $d$ for the \emph{down-arrow}. As before, we use $|c|=w$ to denote the width of a span-based condition $c$, $R^c$ to denote its root, and $p^c_i=(s^c_i,c_i)$ with span $s^c_i=\spanof{u^c_i}{d^c_i}$ its $i$-th branch. Finally, we use $I^c_i$ for the interface of span $s^c_i$ and $P^c_i$ ($=R^{c_i}$) for its target (hence $u^c_i:I^c_i\to R^c$ and $d^c_i:I^c_i\rightarrow P^c_i$). In all these cases, we may omit the superscript $c$ if it is clear from the context. Pictorially, $c$ can be visualised as in \fcite{sb-condition}.
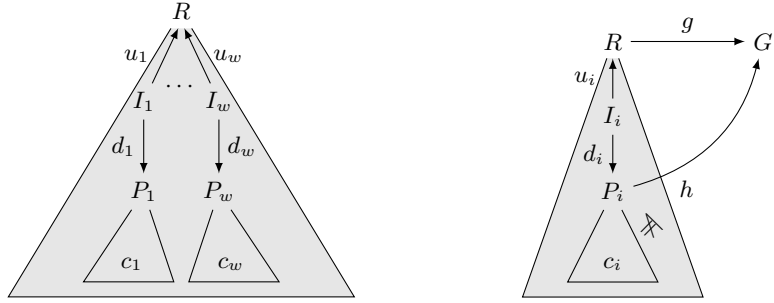
\begin{figure}[t]
\centering
\subcaptionbox
  {Condition $c=(R,p_1\ccdots p_w)$, with $p_i=(\spanof{u_i}{d_i},c_i)$ for $1\leq i\leq w$
   \flabel{sb-condition}}
  [.45\textwidth]
  {\begin{tikzpicture}[on grid]
  \node (R) {$R$};
  \tri{Rc}{3.8}{2.3}{4.6}{}
  \node (I1) [below left=1.2 and .5 of R] {$I_1$};
  \node (P1) [below left=1.2 and .0 of I1] {$P_1$};
  \node [below= of R] {$\cdots$};
  \node (Iw) [below right=1.2 and .5 of R] {$I_w$};
  \node (Pw) [below right=1.2 and .0 of Iw] {$P_w$};

  \tri{P1}{1.2}{.4}{1.2}{$c_1$}
  \tri{Pw}{1.2}{.8}{1.2}{$c_w$}
  
  \path (I1) edge[->] node[left=.1] {$u_1$} (R.260)
        (I1) edge[->] node[left] {$d_1$} (P1)
        (Iw) edge[->] node[right=.1] {$u_w$} (R.280)
        (Iw) edge[->] node[right] {$d_w$} (Pw);
\end{tikzpicture}}
\qquad
\subcaptionbox
  {$g\sat c$, with responsible branch $p_i=(\spanof{u_i}{d_i},c_i)$ and witness $h$ such that $\spanof{u_i}{d_i}\commutes (g,h)$
   \flabel{sb-satisfaction}}
  [.45\textwidth]
  {\begin{tikzpicture}[on grid]
  \node (R) {$R$};
  \tri{R}{3.4}{1.2}{2.4}{}
  \node (Ii) [below=of R] {$I_i$};
  \node (Pi) [below=of Ii] {$P_i$};
  \tri{Pi}{1.2}{.6}{1.2}{$c_i$}
  \node (G) [right=2 of R] {$G$};

  \path (R) edge[->] node[above] {$g$} (G)
        (Ii) edge[->] node[left=.1] {$u_i$} (R)
        (Ii) edge[->] node[left] {$d_i$} (Pi)
        (Pi) edge[->,bend right] node[pos=0.2,below right] (h) {$h$} (G)
        (h) edge[draw=none] node[sloped,allow upside down] {$\nsat$} (Pi-label);
\end{tikzpicture}}
\vspace*{-2mm}
\caption{Visualisations for span-based conditions}
\end{figure}
As we will see in \scite{categories}, we can reconstruct an arrow-based condition from a span-based one.

As examples of span-based conditions, \fcite{sb-conditions} shows counterparts for the three arrow-based conditions in \fcite{ab-conditions}: both $b_1$ and $b_1'$ are equivalent to $c_1$ (which is rather non-obvious in the case of $b_1'$), whereas $b_2$ is equivalent to $c_2$ and $b_3$ to $c_3$.

\begin{figure}[t]
\centering
\begin{tikzpicture}[on grid,node distance=.8 and .7]
  \node[graph] (00) {\onenode{x}};
  \node[left=0 of 00.west] {$b_1'$};
  \node[graph,below=of 00] (I-01) {\twonode{x}{x'}}; 
  \node[graph,below=of I-01] (01) {\oneedge{x}{b}{x'}}; 
  \node[graph,below left=of 01] (I-011) {\onenode{x}};
  \node[graph,below=of I-011] (011) {\oneedge{x}{a}{y}};
  \node[graph,below right=of 01] (I-012) {\onenode{x'}};
  \node[graph,below=of I-012] (012) {\oneedge{x'}{c}{z}};
  
  \path
    (I-01) edge[->] (00)
    (I-01) edge[->] (01)
	(I-011) edge[->] (01)
	(I-011) edge[->] (011)
    (I-012) edge[->] (01)
    (I-012) edge[->] (012);
  
  \node[graph,right=3 of 00] (10) {\onenode{x}};
  \node[left=0 of 10.west] {$b_1$};
  \node[graph,below=of 10] (I-11) {\onenode{x}}; 
  \node[graph,below=of I-11] (11) {\oneloopleft{x}{b}}; 
  \node[graph,below left=of 11] (I-111) {\onenode{x}};
  \node[graph,below=of I-111] (111) {\oneedge{x}{a}{y}};
  \node[graph,below right=of 11] (I-112) {\onenode{x}};
  \node[graph,below=of I-112] (112) {\oneedge{x}{c}{z}};
  
  \path
    (I-11) edge[->] (10)
    (I-11) edge[->] (11)
	(I-111) edge[->] (11)
	(I-111) edge[->] (111)
    (I-112) edge[->] (11)
    (I-112) edge[->] (112);
  
  \node[graph,right=3 of 10] (20) {\onenode{x}};
  \node[left=0 of 20.west] {$b_2$};
  \node[graph,below=of 20] (I-21) {\onenode{x}}; 
  \node[graph,below=of I-21] (21) {\oneedge{x}{b}{y}}; 
  \node[graph,below left=of 21] (I-211) {\onenode{y}};
  \node[graph,below=of I-211] (211) {\oneloop{y}{a}};
  \node[graph,below right=of 21] (I-212) {\onenode{y}};
  \node[graph,below=of I-212] (212) {\oneedge{y}{c}{z}};
  
  \path
    (I-21) edge[->] (20)
    (I-21) edge[->] (21)
	(I-211) edge[->] (21)
	(I-211) edge[->] (211)
    (I-212) edge[->] (21)
    (I-212) edge[->] (212);

  \node[graph,right=3 of 20] (30) {\onenode{x}};
  \node[left=0 of 30.west] {$b_3$};
  \node[graph,below left=of 30] (I-31) {\onenode{x}}; 
  \node[graph,below=of I-31] (31) {\oneloop{x}{a}}; 
  \node[graph,below right=of 30] (I-32) {\onenode{x}}; 
  \node[graph,below=of I-32] (32) {\oneedge{x}{b}{y}}; 
  \node[graph,below=of 32] (I-321) {\onenode{y}}; 
  \node[graph,below=of I-321] (321) {\spangraph{v}{c}{y}{c}{z}}; 
  \node[graph,below=of 321] (I-3211) {\twonode{v}{z}}; 
  \node[graph,below=of I-3211] (3211) {\onenode z}; 

  \path (I-31) edge[->] (30)
		(I-31) edge[->] (31)
        (I-32) edge[->] (30)
		(I-32) edge[->] (32)
		(I-321) edge[->] (32)
		(I-321) edge[->] (321)
		(I-3211) edge[->] (321) 
		(I-3211) edge[->] node[right] {\mapping{v&z}} (3211);

  \path
    ($(10)+(-.2,-4)$) edge[<-,bend left=50,dotted] node[below] {$m$} ($(00)+(0,-4)$)
    ($(10)+(.2,-4)$) edge[<-,bend right=50,dotted] node[below] {$m'$} ($(20)+(-.2,-4)$)
    ($(20)+(.2,-4)$) edge[<-,bend right=50,dotted] node[below] {$m''$} ($(30)+(-.2,-4)$);
	
  \path[morphism]
    (01) edge[medge,->] (11)
	(111) edge[medge,bend left=30,->] (011)
	(112) edge[medge,bend left=30,->] (012)
	;
	
  \path[morphism]
    (21) edge[medge,->] (11)
	(111) edge[medge,bend right=30,->] (211)
	(112) edge[medge,bend right=30,->] (212)
	;
	
  \path[morphism]
    (32) edge[medge,bend left=20,->] (21)
	(212) edge[medge,bend right=30,->] (321)
	;
\end{tikzpicture}
\vspace*{-5mm}
\caption{Span-based conditions corresponding to \fcite{ab-conditions}. The dotted arrows indicate forward-shift morphisms (see \excite{sb-morphisms}), with component mappings in brown.}
\flabel{sb-conditions}
\end{figure}
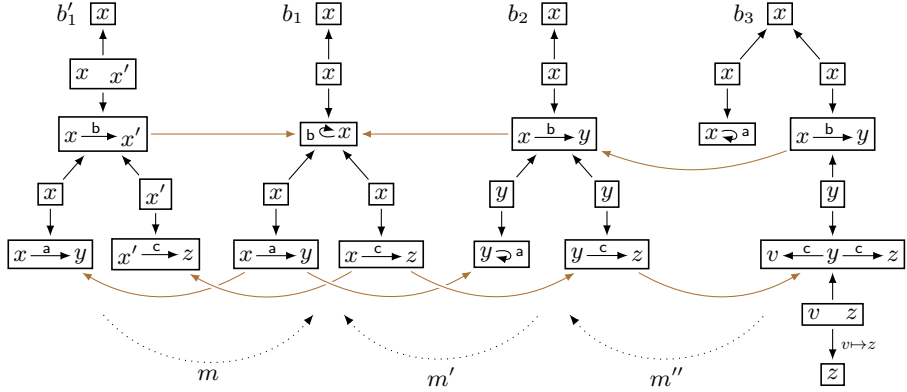

\medskip\noindent The careful reader might have observed that \dcite{sb-condition} is essentially \dcite{ab-condition} instantiated to the category {\SpanC}, thus wondering if we are just repeating the theory of ab-conditions for one specific instance. This is not the case, though, as the definitions that follow (satisfaction, morphisms,\ldots) are phrased using arrows and diagrams of \cat{C}, not of {\SpanC}.\footnote{Technically, this would not fall in our framework because {\SpanC} is not a presheaf topos even if \cat{C} is.}

We now present the modified notion of satisfaction for span-based conditions. First we recognise that the purpose of an arrow $a$ in an ab-condition is essentially to establish ``correct" model/witness pairs $(g,h)$ --- namely, those pairs that commute with $a$ in the sense of satisfying $g=a;h$. We may write $a\commutes (g,h)$ to express this commutation relation. This is what we now formalise for spans, as follows:
\[ \spanof u d \commutes (g,h) \text{ if } u;g=d;h \enspace. \]
Satisfaction of sb-conditions is then defined as follows:

\begin{definition}[satisfaction of span-based conditions]\dlabel{sb-satisfaction}
  Let $c$ be a span-based condition and $g:R\to G$ an arrow from $c$'s root to an object $G$. We say that \emph{$g$ satisfies $c$}, denoted $g\sat c$, if there is a branch $p_i$ and an arrow $h:P_i\to G$ such that
  \begin{enumerate*}
  \item $s_i \commutes (g, h)$, and
  \item $h\nsat c_i$.
  \end{enumerate*}
\end{definition}
Note that this is entirely analogous to \dcite{ab-satisfaction}, especially if we would retrofit the notation $a\commutes (g,h)$ there. Like before, we call $p_i$ the \emph{responsible branch} and $h$ the \emph{witness} of $g\sat c$. Satisfaction of sb-conditions is visualised in \fcite{sb-satisfaction}.

\section{Span-based morphisms}
\slabel{sb-morphisms}

To define morphisms of sb-conditions, we will go over the same ground as for ab-conditions, but take a slightly more abstract view --- as we did for satisfaction by introducing $\commutes$. Having recognised that the \emph{semantics} of a span in sb-conditions is essentially given by the set of model/witness pairs that it commutes with, we can in general consider relations over spans that either preserve or reflect this semantics, in the sense that if $s_1,s_2$ are related, then either $s_1\commutes (g,h)$ implies $s_2\commutes (g,h)$ (preservation) or $s_2\commutes (g,h)$ implies $s_1\commutes (g,h)$ (reflection). Typically, however, the model/witness pairs are not preserved or reflected precisely, but modulo some arrow that gets added to or erased from the model.
\begin{itemize}
\item $(s_1,s_2)$ preserves models adding $v$ if $s_1\commutes (g,h)$ implies $s_2\commutes (v;g,h)$;
\item $(s_1,s_2)$ preserves models erasing $v$ if $s_1\commutes (v;g,h)$ implies $s_2\commutes (g,h)$;
\item $(s_1,s_2)$ reflects models adding $v$ if $s_2\commutes (g,h)$ implies $s_1\commutes (v;g,h)$;
\item $(s_1,s_2)$ reflects models erasing $v$ if $s_2\commutes (v;g,h)$ implies $s_1\commutes (g,h)$.
\end{itemize}
The above is defined on individual pairs. For a relation $\cR$ over spans, we say that $\cR$ preserves [reflects] models adding [erasing] $v$ if it consists of pairs that do so.

\begin{definition}[span source shifters]\dlabel{sb-shifters}
Let us denote by $\cat{C^{sp}}(A,\_)$ the collection of all spans of \cat{C} with source $A$. Let $X,Y$ be objects of \cat{C}. 
\begin{itemize}[topsep=\smallskipamount]
\item A \emph{span source shifter} $\cS$ from $X$ to $Y$ is a target-preserving mapping $\cS \of \cat{C^{sp}}(X,\_) \to \cat{C^{sp}}(Y, \_)$
such that for all spans $s:X\to Z,t:Z\to U$:
\begin{equation}\eqlabel{span shifters are functors}
\cS(s \spcomp t) = \cS(s)\spcomp t \enspace.
\end{equation}
\item Now let $v=A\to B$ be an arrow. $\cS$ is a \emph{forward span source shifter for $v$} if it is a span source shifter from $A$ to $B$ that preserves models erasing $v$, i.e., such that for all spans $s:A\to C$ and all pairs $g:A\to G,h:C \to G$:
\begin{equation}\eqlabel{sb-forward condition}
s\commutes (v;g,h) \enspace\text{ implies }\enspace \cS(s) \commutes (g,h) \enspace.
\end{equation}
A forward span source shifter is called \emph{complete} if it also reflects models adding $v$; i.e., if the ``implies" in \eqref{sb-forward condition} is an ``if and only if".

\item Instead, $\cS$ is a \emph{backward span source shifter for $v$} if it is a span source shifter from $B$ to $A$ that reflects models erasing $v$, i.e., such that for all spans $s:B\to C$ and all pairs $g:B\to G,h:C\to G$:
\begin{equation}\eqlabel{sb-backward condition}
\cS(s) \commutes (v;g,h) \enspace\text{ implies }\enspace s\commutes (g,h) \enspace.
\end{equation}
A backward span source shifter is called \emph{complete} if it also preserves models adding $v$; i.e., if the ``implies" in \eqref{sb-backward condition} is an ``if and only if".
\end{itemize}
\end{definition}
Since span source shifters from $X$ are, just as in the arrow-based case, completely determined by $\cS(\id_X)$ (due to \eqref{span shifters are functors}), we can enumerate the choices of forward and backward shifters for $v$ that are available in general, i.e., without having additional information about $v$.
\begin{description}
\item[Forward shifting.] For $\cS$ to be a forward span source shifter for $v:A\to B$, there must be a span $t=\cS(\id_A):B\to A$, which automatically determines $\cS(s)=t\spcomp s$ for all $s:A\to C$. Without additional information about $v$, there are essentially two options for $t$: $t=\spanof{v}{\id_A}$ or $t=\spanof{\id_B}{x}$ for some section $x$ of $v$, i.e., such that $x;v=\id_B$.

\item[Backward shifting.] For $\cS$ to be a backward span source shifter for $v:A\to B$, there must be a span $t=\cS(\id_B):A\to B$, which automatically determines $\cS(s)=t \spcomp s$ for all $s:B\to C$. Again, there are essentially two options: $t=\spanof{\id_A}{v}$, or $t=\spanof{x}{\id_B}$ for some section $x$ of $v$, i.e., such that $x;v=\id_B$.
\end{description}
This gives rise to the following four candidate shifters for $v:A\to B$ (which still have to be shown to satisfy the conditions of \dcite{sb-shifters}):
\begin{align*}
\dF v & : s \mapsto \spanof{v}{\id_A}\spcomp s \\
\rF v x& : s \mapsto \spanof{\id_B}{x}\spcomp s \text{ where $x;v=\id_B$} \\
\dB v& : s \mapsto \spanof{\id_A}{v}\spcomp s \\
\rB v x& : s \mapsto \spanof{x}{\id_B}\spcomp s \text{ where $x;v=\id_B$}
\end{align*}
These are visualised in \fcite{sb-source shifters} (in a simplified construction that omits the intermediate steps of the span composition where possible). We call $\dF v$ and $\dB v$ \emph{direct} and $\rF v x$ and $\rB v x$ \emph{split}. It turns out that, indeed, each of these four options gives rise to a valid shifter; moreover, the direct ones are complete.
\begin{figure}[t]
\centering
\begin{tikzpicture}[on grid]
  \node (F) {$\cF_v^\circ$};
  \node (A) [below left=.7 and .75 of F] {$A$};
  \node (B) [right=1.5 of A] {$B$};
  \node (I) [below=1.5 of A] {};
  \node (C) [below=1.5 of I] {$C$};

  \path (A) edge[->] node[above] {$v$} (B)
        (I) edge[->] node[left] {$u$} (A) 
        (I) edge[->] node[left] {$d$} node[right] {$d'$} (C)
		(I) edge[->] node [right] {$u'$} (B);
\end{tikzpicture}
\quad
\begin{tikzpicture}[on grid]
  \node (F) {$\cF_v^x$};
  \node (A) [below left=.7 and .75 of F] {$A$};
  \node (B) [right=1.5 of A] {$B$};
  \node (I) [below=1.5 of A] {};
  \node (C) [below=1.5 of I] {$C$};
  \node (B2) [below=1 of B] {$B$};
  \node (I2) [below=1.5 of B2] {};
  
  \path (A) edge[->] node[above] {$v$} (B)
        (I) edge[->] node[left] {$u$} (A) 
        (I) edge[->] node[left] {$d$} (C)
		(B2) edge[->] node[right] {$\id_B$} (B)
		(B2) edge[->] node[below] {$x$} (A)
		(I2) edge[->] node[right] {$u'$} (B2)
		(I2) edge[->] node[below] {$x'$} (I)
		(I2) edge[->] node[below] {$d'$} (C);

  \path	(I2) edge[-{Straight Barb[black,length=5pt,width=10pt]},white] +(-3mm,6mm);
\end{tikzpicture}
\quad
\begin{tikzpicture}[on grid]
  \node (F) {$\cB_v^\circ$};
  \node (A) [below left=.7 and .75 of F] {$A$};
  \node (B) [right=1.5 of A] {$B$};
  \node (I) [below=1.5 of B] {};
  \node (C) [below=1.5 of I] {$C$};
  \node (I2) [below=1.5 of A] {};
  
  \path (A) edge[->>] node[above] {$v$} (B)
        (I) edge[->] node[right] {$u$} (B) 
        (I) edge[->] node[right] {$d$} (C)
		(I2) edge[->>] node[above] {$v'$} (I)
		(I2) edge[->] node[left] {$u'$} (A)
		(I2) edge[->] node[below left] {$d'$} (C);

  \path	(I2) edge[-{Straight Barb[black,length=5pt,width=10pt]},white] +(+3mm,3mm);
\end{tikzpicture}
\quad
\begin{tikzpicture}[on grid]
  \node (F) {$\cB_v^x$};
  \node (A) [below left=.7 and .75 of F] {$A$};
  \node (B) [right=1.5 of A] {$B$};
  \node (B2) [below=of B] {$B$};
  \node (I) [below=of B2] {};
  \node (C) [below=of I] {$C$};
  
  \path (A) edge[->] node[above] {$v$} (B)
        (I) edge[->] node[right] {$u$} (B2) 
        (I) edge[->] node[right] {$d$} node[left] {$d'$} (C)
		(B2) edge[->] node[right] {$\id_B$} (B)
		(B2) edge[->] node[below] {$x$} (A)
		(I) edge[->,bend left] node[below left] {$u'$} (A);
\end{tikzpicture}
\caption{Forward and backward span source shifters. In each case, $\spanof{u'}{d'}$ is the image of $\spanof{u}{d}$.}
\flabel{sb-source shifters}
\end{figure}
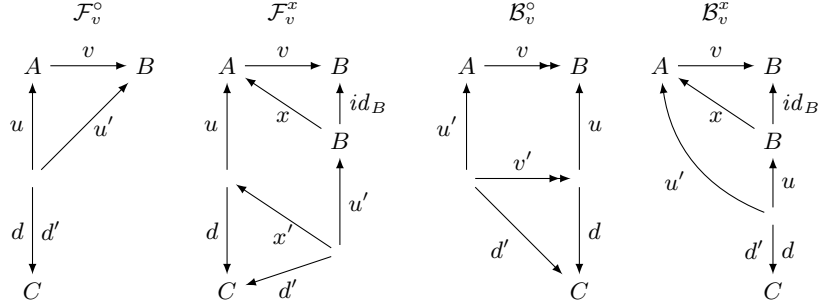
\begin{proposition}[span source shifters]\plabel{sb-source shifters}
\begin{enumerate}[topsep=\smallskipamount]
\item $\cF_v^\circ$ is a complete forward span source shifter for $v$, and $\cF_v^x$ is a forward span shifter for $v$ if $v$ is a split epi with section $x$.
\item $\cB_v^\circ$ is a complete backward span source shifter for $v$ if $v$ is epi, and $\cB_v^x$ is a backward span shifter for $v$ if $v$ is a split epi with section $x$.
\end{enumerate}
\end{proposition}
\iffest
The proof (in \acite{proofs}) can be constructed directly from the diagrams in \fcite{sb-source shifters}.
\fi
\begin{fullorname}
\begin{proof}
In the proof, we make use of the diagrams (especially the arrow names) in \fcite{sb-source shifters}.
\begin{enumerate}
\item We first show that $\cF_v^\circ$ is a complete forward span source shifter. Let $g:B\to G$ and $h:C\to G$; then $\spanof{u}{d}\commutes (g,h)$ and $\spanof{u'}{d'}\commutes (g,h)$ are both equivalent to $u;v;g=d;h$, hence they are equivalent to one another. It follows that $\cF_v^\circ$ both  preserves models erasing $v$ and reflects models adding $v$.

We now show that $\cF_v^x$ is a forward source shifter, i.e., that it preserves models erasing $v$. Let $g:B\to G$ and $h:C\to G$ be such that $\spanof{u}{d}\commutes (v;g,h)$, meaning $u;v;g=d;h$. It follows that $x';u;v;g=x';d;h$ and hence $u';g=u';x;v;g=d';h$, hence $\spanof{u'}{d'}\commutes (g;h)$.

\item We first show that $\cB_v^\circ$ is a complete backward span source shifter for $v$ if $v$ is epi. First note that in toposes the pullback of an epi is epi; hence, as already indicated in the figure, $v'$ is epi. Now let $g:B\to G$ and $h:C\to G$; then $\spanof{u}{d}\commutes (g,h)$ comes down to $u;g=d;h$ whereas $\spanof{u'}{d'}\commutes (v;g,h)$ comes down to $u';v;g=d';h$, which is equivalent to $v';u;g=v';d;g$. Because $v'$ is epi, $u;g=d;h$ if and only if $v';u;g=v';d;g$, implying that $\cB_v^\circ$ both reflects models erasing $v$ and preserves models adding $v$.

We now show that $\cB_v^x$ is a backward span source shifter for $v$, i.e., that it reflects models erasing $v$. Let $g:B\to G$ and $h:C\to G$ be such that $\spanof{u'}{d'}\commutes (v;g,h)$, meaning $u';v;g=d';h$. It follows that $u;g=u;x;v;g=u';v;g=d';h=d;h$, hence $\spanof u d\commutes (g,h)$.\qed
\end{enumerate}
\end{proof}
\end{fullorname}

\medskip\noindent
The step from span source shifters to span \emph{root} shifters is completely identical to the arrow-based case; we omit the definition here. A new aspect is the following.

\begin{proposition}
The trivial root shifter $\cI_{A,B}$ is both a complete forward span root shifter and a complete backward span root shifter for any arrow $v:A\to B$. 
\end{proposition}
Span root shifters satisfy the analogous properties of arrow root shifters formulated in Propositions \pref{ab-root shifters preserve} (model preservation) and \pref{ab-shifters compose} (identities and composition), both extended with the case of complete shifters. 

\begin{proposition}[span root shifters preserve models]\plabel{sb-shifters preserve}
Let $v:A\to B$.
\begin{enumerate}[topsep=\smallskipamount]
\item If $c\in \SC A$ and $\cF$ is a forward span root shifter for $v$ that is defined on $c$, then $v;g\sat c$ implies $g\sat \cF(c)$. If, moreover, $\cF$ is complete then also $g\sat \cF(c)$ implies $v;g\sat c$.
\item If $c\in \SC B$ and $\cB$ is a backward span root shifter for $v$ that is defined on $c$, then $v;g\sat \cB(c)$ implies $g\sat c$. If, moreover, $\cB$ is complete then also $v;g\sat \cB(c)$ implies $g\sat c$.
\end{enumerate}
\end{proposition}
Forward and backward shifters again compose, also for complete ones.
\iffest
The precise statement is given in \pcite{sb-shifters compose}.
\fi

\begin{fullorname}[\pname{sb-shifters compose}]
\begin{proposition}[span shifters compose]\plabel{sb-shifters compose}
\begin{enumerate}[topsep=\smallskipamount]
\item If $\cU$ is a span source [root] shifter from $X$ to $Y$ and $\cV$ a span source [root] shifter from $Y$ to $Z$, then $\cU;\cV$ (understood as partial function composition) is a span source [root] shifter from $X$ to $Z$. 
\item If $\cU$ and $\cV$ are [complete] forward span source [root] shifters for $u: X\to Y$ and $v: Y \to Z$, respectively, then $\cU;\cV$ is a [complete] forward source [root] shifter for $u;v$.
\item Dually, if $\cU$ and $\cV$ are [complete] backward span source [root] shifters for $u: Y\to X$ and $v: Z \to Y$, respectively, then $\cU;\cV$ is a [complete] backward span source [root] shifter for $v;u$.
\end{enumerate}
\end{proposition}
\end{fullorname}
%
%
We call the span source and root shifters of the shape $\dF v$, $\rF v x$, $\dB v$, $\rB v x$ and $\cI_{A,B}$ \emph{elementary}. It is not the case that every span shifter is elementary; e.g., $\dF{v_1};\rF{v_2}{x_2}$ is in general not equal to $\dF{v_1;v_2}$ or to $\rF{v_1;v_2} x$ for any $x$. However, for the remainder of this paper we will restrict ourselves to span shifters that are at least \emph{composed} from elementary ones, in the following sense:

\begin{assumption}\asslabel{sb-shifters} Let $\cS$ be a span root shifter for $v$.
\begin{itemize}[topsep=\smallskipamount]
\item If $\cS$ is a forward shifter, then $\cS = \cS_1;\cdots;\cS_n$ where each $\cS_i$ is an elementary forward shifter for some $v_i$ such that $v=v_1;\cdots;v_n$. If, moreover, $\cS$ is complete then all $\cS_i$ are complete --- in which case either $\cS$ is trivial or $\cS_i=\dF{v_i}$ for all $i$ and $\cS=\dF v$.
\item If $\cS$ is a backward shifter, then $\cS = \cS_1;\cdots;\cS_n$ where each $\cS_i$ is an elementary backward shifter for some $v_i$ such that $v=v_n;\cdots;v_1$. If, moreover, $\cS$ is complete then all $\cS_i$ are complete --- in which case either $\cS$ is trivial or $\cS_i=\dB{v_i}$ for all $i$ and $\cS=\dB v$.
\end{itemize}
\end{assumption}

A final change with respect to the arrow-based case is that we also abstract the commutation condition in \dcite{ab-morphism} by relying instead on a family of \emph{pattern shift relations} --- which in a sense complement the role of root shifters.
\begin{definition}[pattern shift]
Let $v:P_1\to P_2$ be an arrow. \emph{Pattern shift for $v$} is the relation $\cP_v$ over spans including all pairs $(s_1,s_2)$ such that  $s_i=\spanof{u_i:I_i\to A}{d_i:I_i\to P_i}$ for $i=1,2$ (for some $A$), and there is an arrow $k:I_1\to I_2$ for which both halves of the following diagram commute:
\begin{equation}\eqlabel{pattern shift}
\begin{tikzpicture}[on grid,baseline=(I1.center)]
\node (A) {$A$};
\node (I1) [below left=1 and 1.5 of A] {$I_1$};
\node (P1) [below=1.5 of I1] {$P_1$};
\node (I2) [below right=1 and 1.5 of A] {$I_2$};
\node (P2) [below=1.5 of I2] {$P_2$};
\path
  (I1) edge[->] node[above left,inner sep=1] {$u_1$} (A)
  (I1) edge[->] node[left] {$d_1$} (P1)
  (I2) edge[->] node[above right,inner sep=1] {$u_2$} (A)
  (I2) edge[->] node[right] {$d_2$} (P2);
\path[morphism]
  (P1) edge[->] node[above] {$v$} (P2);
\path[red,color=red]
  (I1) edge[->] node[below] {$k$} (I2);
\end{tikzpicture}
\end{equation}
\emph{Conservative pattern shift for $v$} is the subrelation $\ccP_v\subseteq \cP_v$ consisting of span pairs for which, in addition, the lower square of \eqref{pattern shift} is a pushout.
\end{definition}
The following lists some essential properties of pattern shifting.

\begin{proposition}[pattern shift properties]\plabel{pattern shift}
Let $v:P_1\to P_2$ be an arrow, and let $(s_1,s_2)\in \cP_v$.
\begin{enumerate}[topsep=\smallskipamount]
\item\label{pattern-preserves} For all arrows $g,h$, if $s_2\commutes (g,h)$ then $s_1\commutes (g,v;h)$;
\item\label{pattern-transitive} For all arrows $t:P_2\to P_3$, if $(s_2,s_3)\in \cP_t$ then  $(s_1,s_3)\in \cP_{v;t}$;
\item\label{pattern-congruence} For all spans $s$, if $s\spcomp s_1$ and $s \spcomp s_2$ are defined then $(s\spcomp s_1,s \spcomp s_2)\in \cP_v$.
\end{enumerate}
If $(s_1,s_2)\in \ccP_v$, then in addition:
\begin{enumerate}[resume,topsep=\smallskipamount]
\item\label{conservative-reflects} For all arrows $g,h$, if $s_1\commutes (g,h)$ then $h=v;h'$ and $s_2\commutes (g,h')$;
\item\label{conservative-transitive} For all arrows $t:P_2\to P_3$, if $(s_2,s_3)\in \ccP_t$ then $(s_1,s_3)\in \ccP_{v;t}$;
\item\label{conservative-congruence} For all spans $s=\spanof{a}{\id}$, if $s\spcomp s_1$ and $s\spcomp s_2$ are defined then $(s\spcomp s_1,s\spcomp s_2)\in \ccP_v$.
\end{enumerate}
\end{proposition}
\begin{fullorname}
\begin{proof}
Let $v:P_1\to P_2$ and $t:P_2\to P_3$ be arrows, and let $(s_1,s_2)\in \cP_v$. Let $s_i=\spanof{u_i:I_i\to A}{d_i:I_i\to P_i}$ for $i=1,2,3$.
\begin{enumerate}[topsep=\smallskipamount]
\item Assume $s_2\commutes (g,h)$, hence $u_2;g=d_2;h$. It follows that $u_1;g=k;u_2;g=k;d_2;h=d_1;v;h$, hence $s_1\commutes (g,v;h)$.

\item Assume $(s_2,s_3)\in \cP_{t}$; hence there are $k_v$ and $k_t$ making the pentagons of \eqcite{pattern shift} commute for $v$ and $t$, respectively. Then $k_v;k_t$ likewise satisfies the role of $k$ in \eqref{pattern shift} for $v;t$. If, moreover, the lower square of the constituent pentagons (for $v$ and $t$) are pushouts, then by pushout composition so is the lower square of the composed pentagon, which establishes clause \ref{conservative-transitive}.

\item Let $s=\spanof{b:J\to B}{a:J\to A}$. The following figure shows the pentagon obtained for $(s \spcomp s_1,s \spcomp s_2)$.
\begin{center}
\begin{tikzpicture}[on grid]
\node (A) {$A$};
\node (J1) [left=1.5 of A] {};
\node (J2) [right=1.5 of A] {};
\node (I1) [below left=1 and 1.5 of A] {$I_1$};
\node (P1) [below=of I1] {$P_1$};
\node (I2) [below right=1 and 1.5 of R] {$I_2$};
\node (P2) [below=of I2] {$P_2$};
\path
  (I1) edge[->] node[above left,near end,inner sep=1] {$u_1$} (A)
  (I1) edge[->] node[left] {$d_1$} (P1)
  (I2) edge[->] node[above right,near end,inner sep=1] {$u_2$} (A)
  (I2) edge[->] node[right] {$d_2$} (P2);
\path[morphism]
  (P1) edge[->] node[above] {$v$} (P2);
\path
  (I1) edge[->] node[above] {$k$} (I2);

\node (J) [above=1.5 of A] {$J$};
\node (B) [above=of J] {$B$};
\node (J1) [below left=1 and 1.5 of J,inner sep=1pt] {$J_1$};
\node (J2) [below right=1 and 1.5 of J,inner sep=1pt] {$J_2$};
\path (J) edge[->] node[right] {$b$} (B)
      (J) edge[->] node[right] {$a$} (A)
	  (J1) edge[->] node[above left,inner sep=1] {$u'_1$} (J)
	  (J2) edge[->] node[right] {$a_2$} (I2)
	  (J2) edge[->] node[above right,inner sep=1] {$u'_2$} (J)
	  (J1) edge[->] node[left] {$a_1$} (I1);
\path (J1) edge[-{Straight Barb[black,length=6pt,width=17pt]},white] +(5mm,-2.5mm);
\path (J2) edge[-{Straight Barb[black,length=6pt,width=17pt]},white] +(-5mm,-2.5mm);

\path[red,color=red]
  (J1) edge[->] node[near start,below] {$k'$} (J2);
\end{tikzpicture}
\end{center}
Given that $J_2$ is the pullback object of $(a,u_2)$ and $u'_1;a=a_1;u_1=a_1;k;u_2$, the red $k'$ uniquely exists such that $k';u'_2=u'_1$ and $k';a_2=a_1;k$. Hence $k'$ satisfies the properties of the required mediating morphism between $s\spcomp s_1=\spanof{u'_1;b}{a_1;d_1}$ and $s\spcomp s_2=\spanof{u'_2;b}{a_2;d_2}$.
\end{enumerate}
Now let $(s_1,s_2)\in \ccP_v$.
\begin{enumerate}[resume,topsep=\smallskipamount]
\item Assume $s_1\commutes (g,h)$ for $g:A\to G,h:P_1\to G$, hence $u_1;g=d_1;h$. It follows that $k;u_2;g=d_1;h$, hence by the pushout property of the lower square of \eqref{pattern shift} there is a unique $e:P_2\to G$ such that $u_2;g=d_2;e$ and $h=v;e$, implying $s_2\commutes (g,e)$.

\item Shown as part of the proof of clause \ref{pattern-transitive} above.

\item In the case where $s=\spanof{a}{\id_A}$ for some arrow $a$, the composition $s \spcomp s_i$ is simply $\spanof{a;u_i}{d_i}$. The lower square of the pentagon \eqref{pattern shift} for $(s\spcomp s_1,s\spcomp s_2)$ is then the same as that for $(s_1,s_2)$, hence it is a pushout.
\qed
\end{enumerate}
\end{proof}
\end{fullorname}
For the analogy with the arrow-based case, let $\cA_v=\setof{(a_1,a_2)\mid a_2=a_1;v}$; this constitutes a set of ``arrow pattern shift pairs" that play exactly the role captured by clause \ref{pattern-preserves}, but with respect to the satisfaction relation of ab-conditions. The conservative case (which we did not go into for ab-morphisms) would correspond to $\ccA_v=\setof{(a_1,a_2)\in \cA\mid a_1=a_2;x \text{ with } x;v=\id}$.

Using pattern shift, we define morphisms over sb-conditions as follows.
\begin{definition}[span-based (complete) condition morphism]\dlabel{sb-morphism}
  Given two sb-conditions $c,b$, a \emph{forward-shift [backward-shift] sb-morphism} $m:c\to b$ is a pair $(o,(v_1,m_1)\ccdots (v_{|b|},m_{|b|}))$ where
  \begin{itemize}[topsep=\smallskipamount]
  \item $o:[1,|b|]\to [1,|c|]$ is a function from $b$'s branches to $c$'s branches;
  \item for all $1\leq i\leq |b|$, $v_i:P^c_{o(i)}\to P^b_i$ is a arrow from the pattern of $p^c_{o(i)}$ to that of $p^b_i$ such that $(s^c_{o(i)},s^b_i)\in \cP_{v_i}$;
  \item \emph{Forward shift:} for all $1\leq i\leq |b|$, there is a forward root shifter $\cF_i$ for $v_i$ such that $m_i:b_i \to \cF_i(c_{o(i)})$ is a forward-shift morphism;
  \item \emph{Backward shift:} for all $1\leq i\leq |b|$, there is a backward root shifter $\cB_i$ for $v_i$ such that $m_i:\cB(b_i) \to  c_{o(i)}$ is a backward-shift morphism.
  \end{itemize}
  The morphism $m$ is called \emph{direct} if all $\cF_i$ [$\cB_i$] and $m_i$ are direct, and is called \emph{complete} if $o$ is surjective and for all $1\leq i\leq |b|$, $(s^c_{o(i)},s^b_i)\in \ccP_{v_i}$, $\cF_i$ [$\cB_i$] is complete and $m_i$ is complete.
\end{definition}
Note that, by \asscite{sb-shifters}, only direct (forward or backward) span root shifters are complete, hence also only direct sb-morphisms can be complete.

Just as for the arrow-based case, one of the most essential properties of span-based condition morphisms is that they reflect models. In addition, complete span-based condition morphisms also preserve models. The following lifts \pcite{ab-morphisms preserve models} to sb-conditions, and extends it with the case for completeness.

\begin{proposition}[sb-condition morphisms reflect models]
                   \plabel{sb-morphisms preserve models}
Let $c,b \in \SC{R}$ be span-based conditions. If $m:c\to b$ is an sb-morphism, then $g\sat b$ implies $g \sat c$ for all arrows $g:R\to G$. Moreover, if $m$ is complete, then also $g\sat c$ implies $g\sat b$ for all $g$.
\end{proposition}
\begin{fullorname}
\begin{proof}
By induction on the depth of $m$. If $m:c\to b$ is an sb-morphism and $g\sat b$, then let $p^b_i$ be the responsible branch and $h$ the witness such that $\spanof{u^b_i}{d^b_i}\commutes (g,h)$ and $h\nsat b_i$, and let $j=o(i)$. Since $(s^c_j,s^b_i)\in \cP_{v_i}$, it follows by \pcite{pattern shift}.\ref{pattern-preserves} that $\spanof{u^c_j}{d^c_j}\commutes (g,v_i;h)$. Now assume (ad absurdum) that $v_i;h\sat c_j$. This will lead to a contradiction, hence $g\sat c$ with responsible branch $p^c_j$ and witness $v_i;h$.
\begin{itemize}[topsep=\smallskipamount]
\item For the forward-shift case, due to \pref{sb-shifters preserve} it follows that $h\sat \cF_i(c_j)$, and hence by the induction hypothesis the existence of $m_i:b_i \to \cF_i(c_j)$ implies $h\sat b_i$, contradicting the above.

\item For the backward-shift case, by the induction hypothesis the existence of $m_i:\cB_i(b_i)\to c_j$ implies $v_i;h\sat \cB_i(b_i)$, and hence due to \pref{sb-shifters preserve} it follows that $h\sat b_i$, contradicting the above.
\end{itemize}
If $m$ is complete and $g\sat c$, then let $p^c_j$ be the responsible branch and $h$ the witness such that $\spanof{u^c_j}{d^c_j}\commutes (g,h)$ and $h\nsat c_j$. Let $i$ be such that $j=o(i)$ (which exists because $o$ is surjective). Since $(s^c_j,s^b_i)\in \ccP_{v_i}$, it follows by \pcite{pattern shift}.\ref{conservative-reflects} that $\spanof{u^b_i}{d^b_i}\commutes (g,h')$ for some $h'$ such that $h=v_i;h'$. Now assume (ad absurdum) that $h'\sat b_i$. This will lead to a contradiction, hence $g\sat b$ with responsible branch $p^b_i$ and witness $h'$.
\begin{itemize}[topsep=\smallskipamount]
\item For the forward-shift case: by the induction hypothesis, the existence of the complete $m_i:b_i \to \cF_i(c_j)$ implies $h'\sat \cF_i(c_j)$. Since $\cF_i$ is complete, due to \pref{sb-shifters preserve} it follows that $h=v_i;h'\sat c_j$, contradicting the above.
\item For the backward-shift case, since $\cB_i$ is complete, due to \pref{sb-shifters preserve} it follows that $h=v_i;h'\sat \cB_i(b_i)$, and hence by the induction hypothesis the existence of the complete $m_i:\cB_i(b_i)\to c_j$ implies $h\sat c_j$, contradicting the above.
\qed
\end{itemize}
\end{proof}
\end{fullorname}
\begin{example}\exlabel{sb-morphisms}
For the span-based conditions in \fcite{sb-conditions}, we have a complete forward-shift morphism $m:b_1'\to b_1$, providing evidence that $b_1\equiv b_1'$; moreover, there are also morphisms $m':b_2\to b_1$ and $m'':b_3\to b_2$. All the required morphism arrows are already shown in the figure. Note that, in \excite{ab-morphisms}, we observed that there does \emph{not} exist an arrow-based morphism from $c_2$ to $c_1$; hence the existence of a span-based morphism between the equivalent $b_2$ and $b_1$ is concrete evidence for our claim, in \scite{categories} below, that the structure of span-based conditions allows more morphisms between them.
\end{example}
We now once more return to the level of categories, using the exact same constructions for identity morphisms and morphism composition as in the arrow-based case (see \eqref{morphism composition} and \eqref{id-morphism}).
We can therefore lift and extend \thcite{ab-categories} to the span-based case.

\begin{theorem}[categories of span-based conditions]\thlabel{sb-categories}
Consider the following categories having sb-conditions as objects and different arrows: 
$\SBC^\forw$ with forward-shift morphisms, $\SBC^{\back}$ with backward-shift morphisms, $\SBC^{\forw{\circ}}$ with complete forward-shift morphisms, $\SBC^{\entails}$ with the preorder of semantic entailment ($c \leq b$ iff $c\entails b$), and $\SBC^\equiv$ with the preorder (actually, equivalence) of semantic equivalence. These are well-defined, and there are identity-on-objects functors $\op{(\SBC^\forw)} \to \SBC^{\entails}$, $\op{(\SBC^\back)} \to \SBC^{\entails}$ and $\op{(\SBC^{\forw{\circ}})} \to \SBC^{\equiv}$.
\end{theorem}
We end our presentation of sb-conditions by again establishing two indexed categories, now based on the category of span source shifters $\bC^\sshift$, lifting \pcite{ab indexed} to the span-based case.

\begin{proposition}[$\bC^\sshift$-indexed categories]\plabel{sb indexed}
There are functors $\SCx^\forw$ and $\SCx^\back$ from $\op{(\bC^\sshift)}$ to $\Cat$, establishing $\bC^\sshift$-indexed categories.
\end{proposition}

\section{From arrow-based to span-based and back}
\slabel{categories}

Span-based conditions are richer in morphisms than arrow-based conditions, but semantically equivalent. In this subsection we show the existence of faithful, semantics-preserving functors from the categories of forward- and backward-shift arrow-based conditions to their span-based counterparts, and we also show equivalence of the entailment-based categories.

First, we inductively define the ``natural'' span-based conditon $\cN(c)$ for an arbitrary arrow-based condition $c\in \AC R$ and $\cN(p)$ for a branch $p\in \AB R$. 
\begin{align*}
\cN: c = (R,p_1\ccdots p_w) & \mapsto (R,\cN(p_1)\ccdots \cN(p_w)) \\
      p=(a,c) & \mapsto (\spanof{\id_R}{a},\cN(c)) \enspace.
\end{align*}
It follows (by induction) that $\cN(c)\in \SC R$ and $\cN(p)\in \SB R$. This transformation fully preserves the (satisfaction-based) semantics of conditions.

\begin{proposition}[$\cN$ preserves semantics]\plabel{N preserves semantics}
Let $c\in \AC R$. For all $g\of R\to G$, $g\sat c$ iff $g\sat\cN(c)$. As a consequence, for all $b\in \AC R$, $b\entails c$ if and only if $\cN(b)\entails \cN(c)$.
\end{proposition}
This is actually implied by \pcite{A preserves semantics} in combination with \lcite{N-A} below. Morphisms remain essentially identical under $\cN$:
\[ \cN: (o,(v_1,m_1)\ccdots (v_w,m_w)) \mapsto (o,(v_1,\cN(m_1))\ccdots (v_w,\cN(m_w))) \enspace. \]
$\cN$ both preserves and reflects morphisms, in the following sense.
\begin{proposition}
$m\of c\to b$ is a forward-shift [backward-shift] ab-condition morphism if and only if $\cN(m)\of \cN(c)\to \cN(b)$ is a forward-shift [backward-shift] sb-condition morphism.
\end{proposition}
\begin{proof}[sketch]
There are two things to be shown: namely,
\begin{enumerate*}[label=\emph{(\roman*)}]
\item the action of arrow-based forward and backward shifters on $a$ coincides with that of span-based forward and backward shifters on $\spanof{id}{a}$, and
\item the commutation condition $a_1=v;a_2$ (for ab-condition morphisms) holds if and only if $(\spanof{\id}{a_1},\spanof{\id}{a_2})\in \cP_v$.
\end{enumerate*}
The latter is immediate. For the former, briefly using $\hat \cS$ to denote span shifters, the essential property is that $\hat\cB_v^\circ(\spanof{\id}{a})=\spanof{id}{\cB_v(a)}$ and $\hat\cF_v^s(\spanof{\id}{a})=\spanof{id}{\cF_v^s(a)}$. (Span shifters of the form $\hat\cB_v^x$ or $\hat\cF_v^\circ$ do not preserve the property that the left span-leg is the identity, hence are not suitable to support morphisms from $\cN(c)$ to $\cN(b)$.)
\qed
\end{proof}
Since identities and morphism composition are obviously preserved by $\cN$, we have the following result, also illustrated in \fcite{categories}:

\begin{theorem}[arrow-based to span-based functors]\thlabel{ab-to-sb functors}
$\cN$ is a faithful functor from $\ABC^\forw$ to $\SBC^\forw$, from $\ABC^\back$ to $\SBC^\back$ and (with the appropriate, obvious mapping of arrows) from $\ABC^{\entails}$ to $\SBC^{\entails}$, which commute with the functors of \thcite{ab-categories} and \thcite{sb-categories}.
\end{theorem}
From span-based to arrow-based conditions with forward- or backward-shift morphisms, there does not exist a functor, as the span-based categories are richer in morphisms --- for instance, \fcite{sb-conditions} shows an example span-based morphism $m':b_2\to b_1$, whereas no arrow-based morphism exists between the equivalent $c_2$ and $c_1$ in \fcite{ab-conditions} (confer Examples \exref{ab-morphisms} and~\exref{sb-morphisms}). However, we now show that span-based conditions themselves are semantically not more expressive than arrow-based ones: there does exist a semantics-preserving mapping from $\SC A$ to $\AC A$ for all $A$.

Let $(d'_s,u'_s)$ denote the pushout cospan for a span $s=\spanof u d$. We inductively define $\cA$ on sb-conditions and sb-branches, as follows:
\begin{align*}
\cA: (R,p_1\ccdots p_w) & \mapsto (R,\cA(p_1)\ccdots \cA(p_w)) \\
     (s,c) & \mapsto (d'_s,\cA(\dF{u'_s}(c))) \enspace.
\end{align*}
For instance, taking again the ab-conditions in \fcite{ab-conditions} and their sb-condition counterparts in \fcite{sb-conditions}, it can be checked that $\cA(b_1)=\cA(b'_1)=c_1$, $\cA(b_2)=c_2$ and $\cA(b_3)=c_3$.

$\cA$ acts as the (left) inverse of the action of $\cN$ on objects, due to $d'_s=\id$ and $u'_s=a$ for $s=\spanof{\id}{a}$:
\begin{lemma}\llabel{N-A}
For any arrow-based condition $c$, $\cA(\cN(c))=c$.
\end{lemma}
The following property is the counterpart of \pcite{N preserves semantics}, and actually implies it thanks to \lcite{N-A}:
\begin{proposition}[$\cA$ preserves semantics]\plabel{A preserves semantics}
If $c\in \SC R$, then $\cA(c)$ is an arrow-based condition in $\AC R$ such that for all $g\of R\to G$, $g\sat c$ iff $g\sat\cA(c)$. In consequence, for all $b\in \SC R$, $b\entails c$ if and only if $\cA(b)\entails \cA(c)$.
\end{proposition}
\begin{proof}[sketch]
By induction on the depth of $c$, using that $g=d'_s;h$ iff $s\commutes (g,u'_s;h)$ for any model/witness pair $(g,h)$, in combination with the fact that $\dF{u'_s}$ preserves models adding $u'_s$ and reflects them erasing $u'_s$ (\pcite{sb-source shifters}).\qed
\end{proof}
This implies that (as claimed above) span-based conditions are semantically not more expressive than arrow-based categories. The formal statement comes down to an equivalence of the respective categories, also illustrated in \fcite{categories}.
\begin{theorem}[equivalence of arrow-based and span-based entailment]\thlabel{sb-to-ab functor}
$\cA$ is a (full and faithful) functor from $\SBC^{\entails}$ to $\ABC^{\entails}$, surjective on objects, which (together with its left and right adjoint $\cN$) establishes that the categories are equivalent.
\end{theorem}

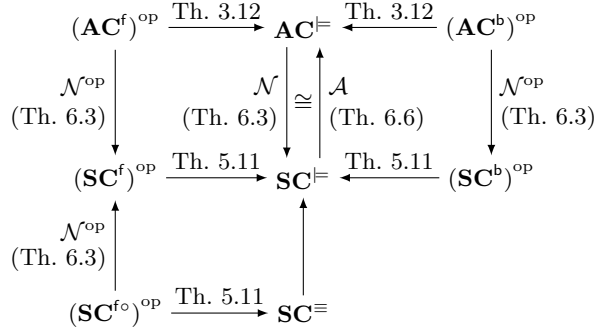
\begin{figure}[t]
\centering
\begin{tikzpicture}[on grid,node distance=1.8 and 2.5]
\renewcommand{\thcite}[1]{Th.~\thref{#1}}
\node (AC-entails) {$\ABC^{\entails}$};
\node[below=2 of AC-entails] (SC-entails) {$\SBC^{\entails}$};
\node[below=of SC-entails] (SC-equiv) {$\SBC^{\equiv}$};
\node[left=of AC-entails] (AC-f) {$\op{(\ABC^\forw)}$};
\node[left=of SC-entails] (SC-f) {$\op{(\SBC^\forw)}$};
\node[left=of SC-equiv] (SC-fo) {$\op{(\SBC^{\forw\circ})}$};
\node[right=of AC-entails] (AC-b) {$\op{(\ABC^\back)}$};
\node[right=of SC-entails] (SC-b) {$\op{(\SBC^\back)}$};

\path
  (AC-f) edge[->] node[above] {\thcite{ab-categories}} (AC-entails)
  (AC-b) edge[->] node[above] {\thcite{ab-categories}} (AC-entails)
  (SC-f) edge[->] node[above] {\thcite{sb-categories}} (SC-entails)
  (SC-b) edge[->] node[above] {\thcite{sb-categories}} (SC-entails)
  (SC-fo) edge[->] node[above] {\thcite{sb-categories}} (SC-equiv)
  (AC-f) edge[->]  node[left,pos=.35] {$\op{\cN}$}
                   node[left,pos=.65] {(\thcite{ab-to-sb functors})}  (SC-f)
  (SC-fo) edge[->] node[left,pos=.65] {$\op{\cN}$}
                   node[left,pos=.35] {(\thcite{ab-to-sb functors})}  (SC-f)
  (AC-entails.230) edge[->]
                   node (N) {}
                   node[left,pos=.35] {$\cN$}
                   node[left,pos=.65] {(\thcite{ab-to-sb functors})} (SC-entails.130)
  (SC-entails.50) edge[->] 
                   node (A) {}
                   node[right,pos=.65] {$\cA$} 
                   node[right,pos=.35] {(\thcite{sb-to-ab functor})} (AC-entails.310)
  (SC-equiv) edge[->] (SC-entails)
  (AC-b) edge[->]  node[right,pos=.35] {$\op{\cN}$}
                   node[right,pos=.65] {(\thcite{ab-to-sb functors})} (SC-b);
				   
\path
  (A) edge[none] node {$\cong$} (N);
\end{tikzpicture}
\caption{Overview of categories and functors studied in this paper}
\flabel{categories}
\end{figure}

\iffull
\input{entailment} 
\fi
\section{Conclusion and future work}
\slabel{conclusion}

The main contributions of this paper are the following:

\begin{itemize}
\item Starting with the existing notion of nested conditions (which we call \emph{arrow-based} to contrast them with a variant developed in the paper), we define structural morphisms that provide evidence for (i.e., ``explain") entailment between conditions. In doing so, we find that there are two independent variants, called forward-shift and backward-shift, that explain different fragments of entailment (\dcite{ab-morphism} and \excite{forward vs backward}).

\item This gives rise to functors from the categories of nested conditions with (forward-shift or backward-shift) morphisms to the preorder of entailment over the same conditions (\thcite{ab-categories}).

\item Observing that there actually exist far fewer morphisms than one would like (relatively few cases of entailment are explained by the existence of morphisms), and suspecting that this is a consequence of redundancy within the structure of nested conditions, we define \emph{span-based} nested conditions in which such redundancy can be avoided (\dcite{sb-condition}).

\item Forward- and backward-shift morphisms are lifted to span-based conditions (\dcite{sb-morphism}), leading again to categories with functors to the preorder of entailment (\thcite{sb-categories}). Moreover, we also characterise \emph{complete} span-based morphisms that in fact imply equivalence of, and not just entailment between, their source and target conditions.

\item There is a faithful embedding of arrow-based conditions (with backward-shift and forward-shift morphisms) into span-based ones, and span-based conditions indeed explain a larger fragment of entailment. However, under the preorder of entailment the two categories are equivalent, meaning that span-based conditions are themselves not more expressive than arrow-based ones (\thcite{sb-to-ab functor}).
\end{itemize}
Besides these main contributions, there are numerous smaller ones, among which are the (admittedly rather technical) existence of indexed categories based on the concept of \emph{shifter} (Props.\ \pref{ab indexed} and \pref{sb indexed}). This is closely related to the notion of shifting in the existing theory of nested conditions (see below).

All investigations in this paper assume a base category that is a presheaf topos, which includes the case of edge-labelled graphs that is used for all the examples (and from which we derive our intuitions). The main results are visualised in \fcite{categories}.

\paragraph{Related work.}

In his work on Existential Graphs~\cite{roberts1973-the-existential-graphs-of-charles-s.-peirce}, Charles S. Peirce proposed a graphical representation of full FOL, equipped with some kinds of graph manipulations which represent sound deductions. Even if equally expressive, comparing Existential Graphs with Nested Conditions looks difficult, because the former are not formalized in a familiar algebraic/categorical way.

Bonchi et.al.~\cite{DBLP:conf/csl/BonchiSS18} have enriched the correspondence between graphs and conjunctive queries of~\cite{DBLP:conf/stoc/ChandraM77}, summarized in the Introduction,  by identifying a common rich categorical structure: \emph{cartesian bicategories}. They introduce \emph{graphical conjunctive queries} as suitable \emph{string diagrams}, i.e.~arrows of a specific free cartesian bicategory, showing that they are as expressive as standard conjunctive queries and, more interstingly, that the freely generated preorder among them is exactly the entailment preorder among queries. Furthermore, exactly the same algebraic structure is shown to arise by considering as arrows cospans of hypergraphs and as preorder the existence of a morphism. This is summarized by a triangular relationship including logical structures (queries), combinatorial structures (hypergraphs), and categorical ones (free cartesian bicategories).  The characterization of conjunctive formulas as arrows of a free cartesian bicategory has been generalized in~\cite{DBLP:journals/corr/abs-2404-18795} to full FOL, but lacking the combinatorial/graphical counterpart. As a possible development of the results of this paper, by equipping conditions with suitable interfaces (encoding free variables) we intend to study the algebraic structure of nested conditions, possibly identifying a suitable cartesian bicategory. If succesful, this could provide the third missing structure (the combinatorial one), allowing to lift to full FOL the triangular correspondence presented in \cite{DBLP:conf/csl/BonchiSS18} for the $\exists$-fragment.

The operation of shifting a condition along an arrow has been exploited for cospan-based conditions in~\cite{bchk:conditional-reactive-systems} to compute weakest pre-conditions and strongest post-conditions for graph transformation systems. This operation, similarly to our notion of shifter, defines a functor between categories of conditions that is shown to have both left and right adjoints, corresponding to a form of existential and universal quantification.  We intend to define logical operations on span-based conditions, and to explore whether the adjunction results still hold.

In~\cite{bchk:conditional-reactive-systems,sksclo:coinductive-techniques-for-satisfiability} nested conditions are defined over an arbitrary category $\bC$, which allows to instantiate the framework beyond presheaf toposes: for example, to the category of graphs and injective morphisms, or to the category of left-linear cospans of an adhesive category. We intend to explore to what extent our assumptions on category $\bC$ can be relaxed in order to be able to apply our results to other, more general settings.

Some papers~\cite{lo:tableau-graph-properties,slo:model-generation,sksclo:coinductive-techniques-for-satisfiability} address the problem of (semi-)deciding satisfiability of nested conditions, by resorting to tableau-based techniques inspired by those of FOL. The proposed algorithms are also able to generate finite models if a formula has one.
It would be interesting to explore if passing from arrow-based to span-based conditions can have an impact on the complexity of proofs, and if morphisms among conditions could allow to relate tableau-based proofs for different formulas.

\paragraph{Future work.}

We see the results of this paper as providing only a start for the study into span-based conditions, giving rise to many natural follow-up questions. Some of those arise in the context of related work and were already discussed above; here are a few more.
\begin{itemize}
\item Is there any independent characterisation of the fragment of entailment that is explained by span-based condition morphisms? Do forward-shift and backward-shift morphisms explain distinct fragments as they do in the arrow-based case?

\item There are many syntactically different (but semantically equivalent) span-based representations for the same property. For instance, it can be shown that replacing any span by another with the same pushout gives rise to an equivalent condition. Is there a useful normal form for span-based conditions, preferably such that, if a morphism exists between two conditions, one also exists between their normal forms?
%
\end{itemize}

\paragraph{Acknowledgements.}
We would like to thank Filippo Bonchi for providing the original inspiration for this work, and together with Nicolas Behr and Barbara König for providing comments in earlier stages.

\bibliographystyle{splncs03}
\bibliography{references.bib}

\iffest
\appendix
\section{Proofs}
\alabel{proofs}

\subsection{Proofs for \scite{ab-morphisms}}

\displayhere{\lname{ab-source shifters}}

\displayhere{\pname{ab-source shifters}}

\displayhere{\lname{ab-root shifter}}

\displayhere{\pname{ab-root shifters preserve}}

\displayhere{\pname{ab-shifters compose}}

\displayhere{\pname{ab-morphisms preserve models}}

\displayhere{\lname{ab-root shifters preserve morphisms}}

\displayhere{\lname{ab-morphisms compose}}

\subsection{Proofs for \scite{sb-conditions}}

\displayhere{\pname{sb-source shifters}}

\displayhere{\pname{sb-shifters compose}}

\displayhere{\pname{pattern shift}}

\displayhere{\pname{sb-morphisms preserve models}}

\fi
\end{document}